\documentclass[
preprint,showpacs,
amsmath,amssymb,aps
]{revtex4-1}

\usepackage{graphicx}
\usepackage{dcolumn}
\usepackage{bm}
\newcommand{\md}{\mathrm d}

\begin{document}


\title{Regularization dependence on the Schwinger--Dyson
equation \\
in Abelian gauge theory:
4D vs 3D cutoff regularization}

\author{Hiroaki Kohyama}
\affiliation{Department of Physics,
National Taiwan University, Taipei 10617, Taiwan.
}

\date{\today}

\begin{abstract}
We study the regularization dependence on the quenched
Schwinger--Dyson equations in general gauge by applying the two
types of regularizations, the four and three dimensional momentum
cutoffs. The obtained results indicate that the solutions are not
drastically affected by the choice of two different cutoff prescriptions.
We then think that both the regularizations can nicely be adopted in
the analyses for the Schwinger--Dyson equations.
\end{abstract}

\pacs{11.15.-q, 11.30.Rd, 12.20.-m}

\maketitle

\section{\label{sec:intro}
Introduction}
The chiral symmetry breaking in the strongly coupling system is
interesting phenomena in quantum chromodynamics. Although the
coupling strength is small in realistic quantum electrodynamics
(QED), the chiral symmetry can be broken when we consider
strongly coupled QED. Then it is interesting to study such a
system in Abelian gauge theories.

For the investigation on the above mentioned chiral symmetry
breaking coming from dynamical mass generation, the analysis based
on the Schwinger--Dyson equation (SDE)  may be an appropriate
approach~\cite{Dyson:1949ha}. The SDE is the set of equations for
green functions whose solutions give the information on the field
renormalization and  dynamical mass generation (for reviews,
see,~\cite{Roberts:1994dr, Roberts:2000aa, Holl:2006ni}).
The analysis of the SDE should conceptually lead the gauge and
regularization independent solutions, since the equations stem from
a renormalizable gauge theory. However, the solutions practically
depend on the chosen gauge, because the equations are derived
through applying some approximations in the intermediate steps.
The extensive analyses on the gauge dependence are given
in~\cite{Kizilersu:2014ela} where the effects of generalised vertices
are studied in unquenched QED with four dimensional cutoff method.
Concerning on the similar generalisations of the equations, a lot of
works have been done with various approaches, 
see, e.g.,~\cite{Oliensis:1990sg, Curtis:1990zs}.
Also, it is known that the physical predictions as well depend on
regularization procedures~\cite{Schreiber:1998ht}, since the quenched
SDE can be regarded as the generalised version of the Nambu Jona
Lasinio-type gap equations in which regularization has effects on
the model predictions~\cite{Inagaki:2015lma}. We then think it may
be interesting to study the regularization dependence on the solutions
of the SDE.

In this letter, we shall numerically solve the SDE in general gauge
with two regularization procedures, the four dimensional (4D) and
three dimensional (3D) cutoff regularizations, then make the
comparison between these two methods. The importance of studying
the equations with the 3D cutoff regularization lies on the fact that the
solutions can smoothly be continued to the ones obtained in the finite
temperature system, since the equations are usually investigated by
using the 3D cutoff scheme at finite temperature~\cite{Fukazawa:1999aj}.

This paper is organised as follows;  Section~\ref{sec:equation}
presents two types of equations. We show the numerical results for
the field renormalization factor and  the dynamically generated mass
in Sec.~\ref{sec:re_ns}. The concluding remarks are given in
Sec.~\ref{sec:conclusion}.

\section{\label{sec:equation}
Schwinger--Dyson equation}
The Schwinger--Dyson equation for fermion self-energy
$\Sigma(P)$ is written by
\begin{equation}
  \Sigma (P)
  = i e^2 \int \frac{{\mathrm d}^4 Q}{(2\pi)^4}
     \gamma^{\mu} D_{\mu\nu}(P-Q) S(Q) \Gamma^{\nu}(P,Q) ,
\end{equation}
where $e$ is the coupling strength, $D_{\mu \nu}(P-Q)$ and $S(Q)$
are the gauge boson and fermion propagators, and
$\Gamma^\nu (P,Q)$ is the vertex function on the gauge boson
and the fermion.
In this letter, we use the capital letters as $P$, $Q$ for expressing the
four dimensional momenta, namely, $P_\mu=(p_0, {\bf p})$ and
$Q_\mu=(q_0,{\bf q})$.
For $D_{\mu \nu}$ and $\Gamma^\nu$, we employ the following
tree-level forms
\begin{eqnarray}
  &&D_{\mu\nu}(K)
  =  \frac{-g_{\mu \nu}}{K^2}
    +(1-\xi) \frac{K_{\mu}K_{\nu}}{K^4},
  \label{eq:photon_prop} \\
  &&\Gamma^{\nu}(P,Q) = \gamma^{\nu},
\end{eqnarray}
with the gauge parameter $\xi$ and $K_\mu = P_\mu -Q_\mu$.

\subsection{\label{subsec:4dcut}
Equations with the four dimensional cutoff}
In the SDE with four dimensional cutoff, we define the fermion
propagator by
\begin{equation}
  S(Q) = \frac{1}
  {A(Q) Q_\mu \gamma^\mu -B(Q)},
\end{equation}
where $A(Q)$ and $B(Q)$ indicate the field strength factor and the
mass function. The insertion of these quantities leads
\begin{align}
 A(P) 
 & P^\mu \gamma_\mu  -B(P) 
 = \nonumber \\
 & P^\mu \gamma_\mu - m_0
   + e^2
   \int \frac{{\mathrm d}^4 Q}{i(2\pi)^4}
   \gamma^{\mu}
     \left[ \frac{-g_{\mu \nu}}{K^2}
          + (1-\xi)\frac{K_{\mu}K_{\nu}}{K^4}
     \right]
     \left(
        \frac{1}{A(Q) Q_\rho \gamma^\rho  -B(Q)}
     \right)
   \gamma^{\nu},
\label{eq:SDE_form}
\end{align}
with the fermion bare mass $m_0$ appearing in the Lagrangian.
Taking the trace after multiplying $P^\rho \gamma_\rho$ and
without the multiplication give the equations for $A(P)$ and $B(P)$
as
\begin{align}
&A(P) = 1
   - \frac{e^2}{P^2}
     \int \frac{ {\mathrm d}^4 Q}{i(2\pi)^4} \,
     \left[
           \xi_{+} \frac{P \cdot Q}{K^2} 
       +2\xi_{-} \frac{(P \cdot K) (Q \cdot K)}{K^4} 
     \right]
     \Delta^{\prime}(Q)
     A(Q), \\
&B(P) = m_0
   - e^2
     \int \frac{{\mathrm d}^4 Q}{i(2\pi)^4} \,
     \left[  \xi_{3} \frac{1}{K^2} 
     \right]
     \Delta^{\prime} (Q)
     B(Q),
\end{align}
where $\xi_{\pm} \equiv 1 \pm \xi$, $\xi_3 \equiv 3+\xi$ and
\begin{align}
  \Delta^{\prime} (Q) = \frac{-1}{A^2(Q)Q^2 -B^2(Q)}.
\end{align}
Performing the angular integration after the Wick rotation, we
obtain the following equations,
\begin{align}
&A(P) = 1 
   +\frac{\alpha \xi}{4\pi} 
     \int_{\delta_{\rm 4D}^2}^{\Lambda_{\rm 4D}^{2}} \md Q^2 
     \left[
        \frac{Q^4}{P^4} \theta(P-Q) + \theta(Q-P)
     \right] 
     \Delta(Q)  A(Q), 
     \label{eq:original_A} \\
&B(P) = m_0
   +\frac{\alpha \xi_3}{4\pi}  
     \int_{\delta_{\rm 4D}^2}^{\Lambda_{\rm 4D}^{2}} \md Q^2  
     \left[
        \frac{Q^2}{P^2} \theta(P-Q) + \theta(Q-P)
     \right]  
     \Delta(Q)  B(Q),
     \label{eq:original_B}
\end{align}
with
\begin{align}
  \Delta(Q) = \frac{1}{A^2(Q)Q^2 +B^2(Q)}.
\end{align}
where $\alpha =e^2/(4\pi)$ and we introduce the ultraviolet and
infrared cutoffs, $\Lambda_{\rm 4D}$ and $\delta_{\rm 4D}$.
These are the equations with the four dimensional cutoff scheme.

\subsection{\label{subsec:3dcut}
Equations with the three dimensional cutoff}
In the three dimensional cutoff, we need to consider the following
fermion propagator 
\begin{equation}
  S(p_0, p) = \frac{1}
  { C(p_0,p)\gamma_0 p^0 + A(p_0,p)\gamma_i p^i -B(p_0,p)}
\end{equation}
with $p = |{\bf p}|$,
since the time and space directions should be treated separately.
After the Wick rotation and a bit of algebras one obtains the
following forms
\begin{align}
 &C(p_0, p) = 1
   + \frac{\alpha}{2\pi^2}
  \int_{-\infty}^{\infty} \md q_{0}
  \int_{\delta_{\rm 3D}}^{\Lambda_{\rm 3D}} \md q \,
     \left[ 
        {\mathcal I}_{CA} A(q_0, q) + {\mathcal I}_{CC} C(q_0, q)
     \right]      \Delta(q_0, q) , \\
&A(p_0, p) = 1
   + \frac{\alpha}{2\pi^2}
  \int_{-\infty}^{\infty} \md q_{0}
  \int_{\delta_{\rm 3D}}^{\Lambda_{\rm 3D}} \md q \,
     \left[
        {\mathcal I}_{AA} A(q_0, q) + {\mathcal I}_{AC} C(q_0, q)
     \right]      \Delta(q_0, q), \\
&B(p_0, p) = m_0
   +  \frac{\alpha}{2\pi^2}
  \int_{-\infty}^{\infty} \md q_{0}
  \int_{\delta_{\rm 3D}}^{\Lambda_{\rm 3D}} \md q \,
     \left[ {\mathcal I}_{B} B(q_0, q)
     \right]      \Delta(q_0, q),
\end{align}
with 
\begin{align}
  \Delta(q_0, q)
  =  \frac{1}{ C^2(q_0, q) q_{0}^2  
                 + A^2(q_0, q) q^2 
                 + B^2(q_0, q)}.
\end{align}
and
\begin{align}
  & {\mathcal I}_{CA} = 
       \xi_{-} \frac{k_0}{p_0}  I_1
     +\xi_{-} \frac{k_0}{p_0}  
        \left[ k_0^2 - q^2 + p^2 \right] I_2, \\
  &  {\mathcal I}_{CC} = 
     -\xi_{+} \frac{q_0}{p_0}  I_1
     +2 \xi_{-} \frac{q_0}{p_0} k_0^2 I_2, \\
  & {\mathcal I}_{AA} = 
      -\frac{2q^2}{p^2}  
      -\frac{1}{2p^2} \left[ \xi_3 k_0^2 + \xi_+(q^2+p^2) \right] I_1
      -\frac{1}{4p^2} \xi_{-} 
        \left[ k_0^4 - (q^2-p^2)^2 \right] I_2, \\
  & {\mathcal I}_{AC} = 
      -\xi_{-} \frac{q_0 k_0}{p^2}  I_1
      -\xi_{-} \frac{q_0 k_0}{p^2}  
        \left[ k_0^2 + q^2 - p^2 \right] I_2, \\
   &  {\mathcal I}_{B} = \xi_3 I_1 , \\
   &I_1 = \frac{q}{2p}
         \ln \frac{k_{0}^2+(q-p)^2}{k_{0}^2+(q+p)^2}, \\
  &I_2 = \frac{q}{2p}
        \left[
          \frac{1}{k_{0}^2+(q-p)^2}-\frac{1}{k_{0}^2+(q+p)^2}
        \right],
\end{align}
where $k_0 = p_0 -q_0$. Thus we need to consider the three unknown
functions $C$, $A$ and $B$ with two variables $p_0$ and $p$, then
the numerical analyses become much more difficult comparing
to the above mentioned four dimensional cutoff case.

\section{\label{sec:re_ns}
Numerical solution}
In this section, we numerically solve the equations with the four and
three dimensional cutoff procedures by using the iteration method,
then make the comparison on the obtained results.

\subsection{\label{subsec:re_4d}
Solutions with the 4D cutoff scheme}
\begin{figure}[!h]
\begin{center}
  \includegraphics[width=6.7cm,keepaspectratio]{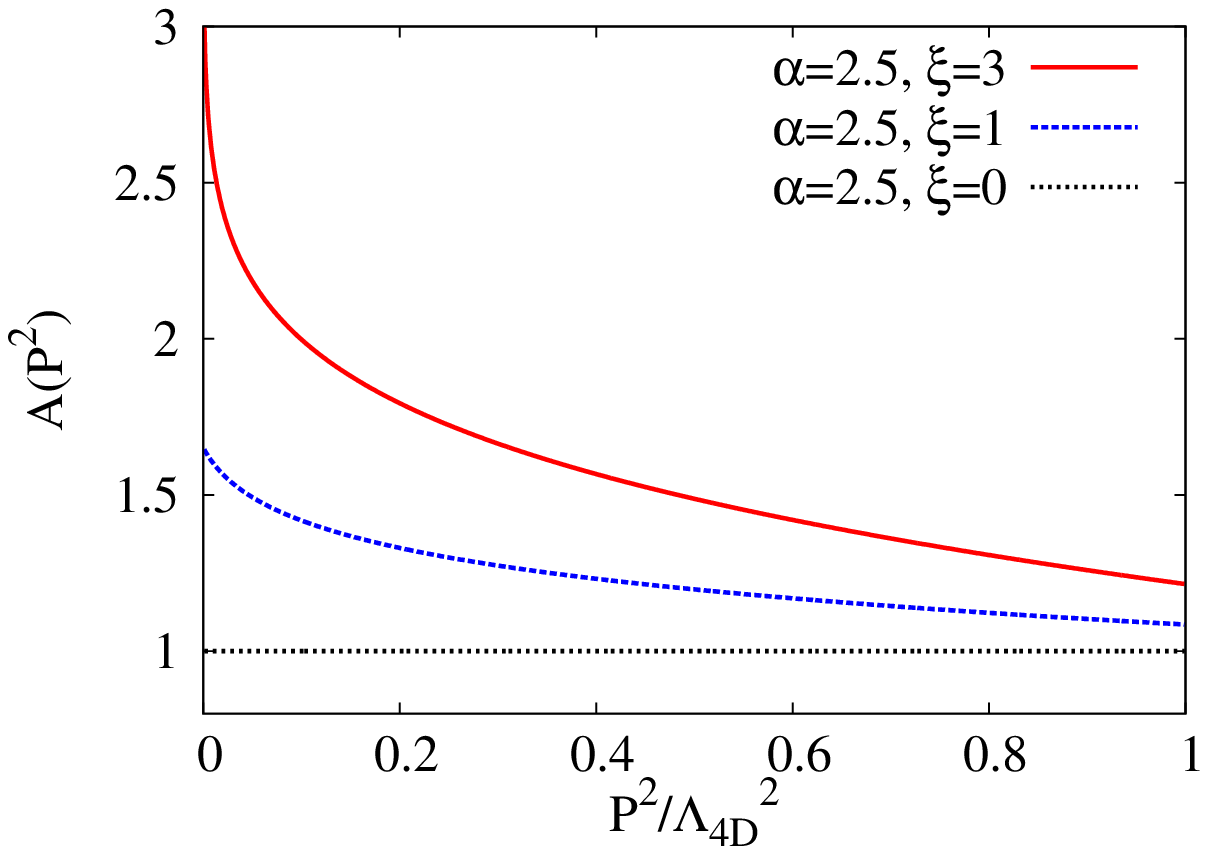}
  \includegraphics[width=6.7cm,keepaspectratio]{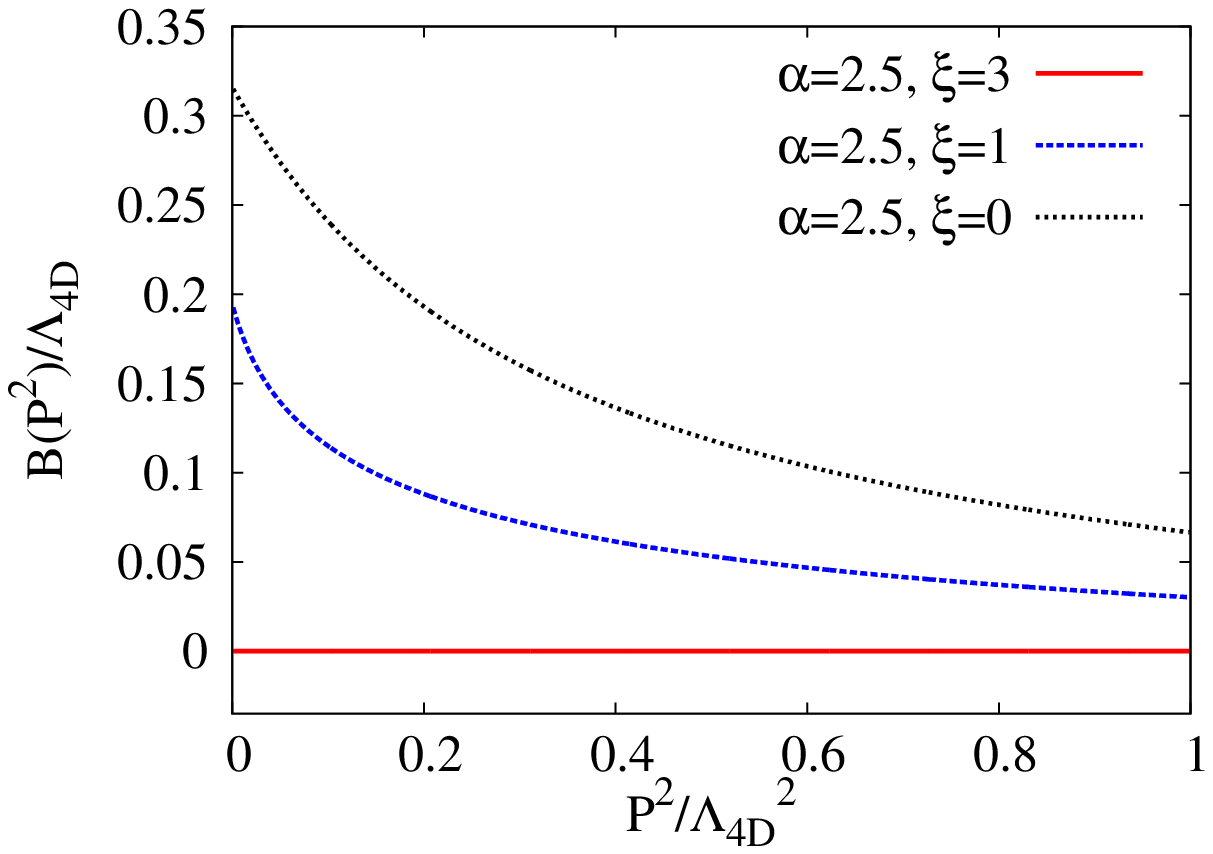}
  \includegraphics[width=6.7cm,keepaspectratio]{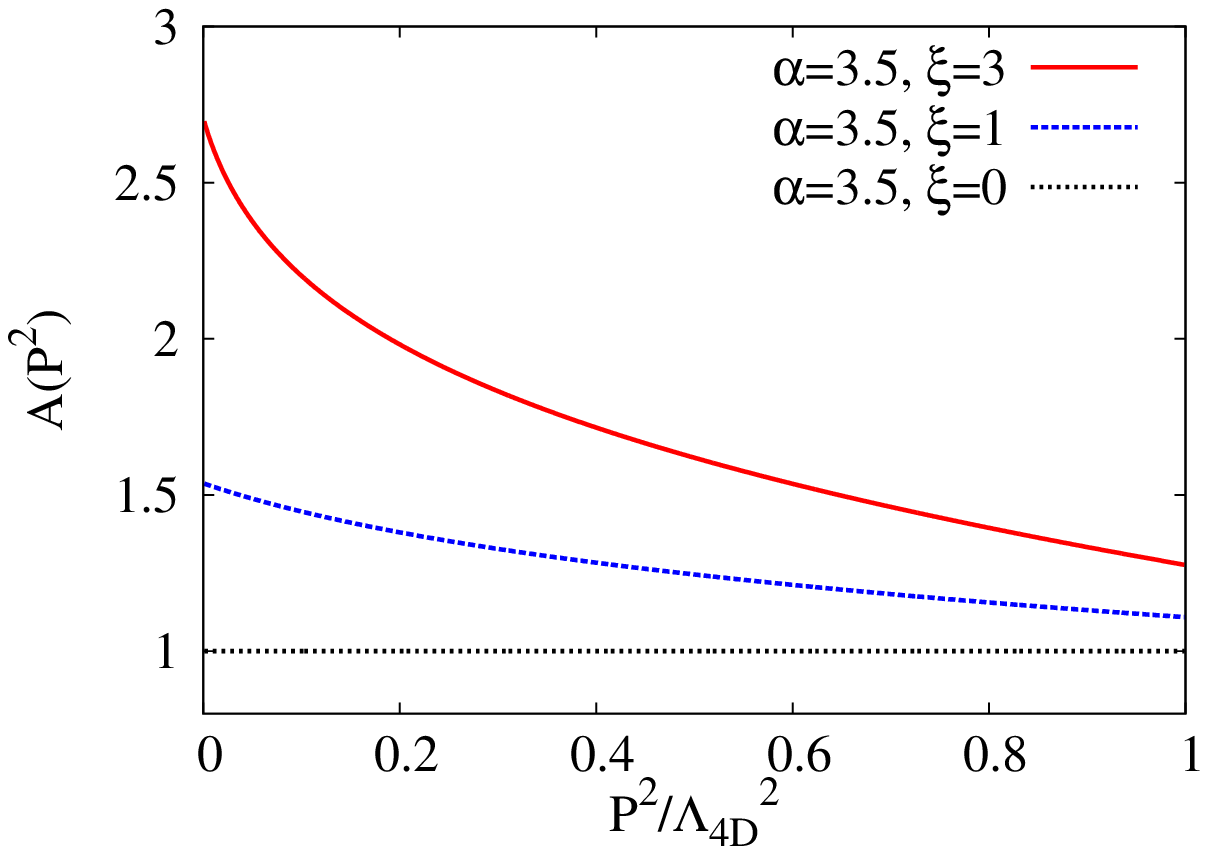}
  \includegraphics[width=6.7cm,keepaspectratio]{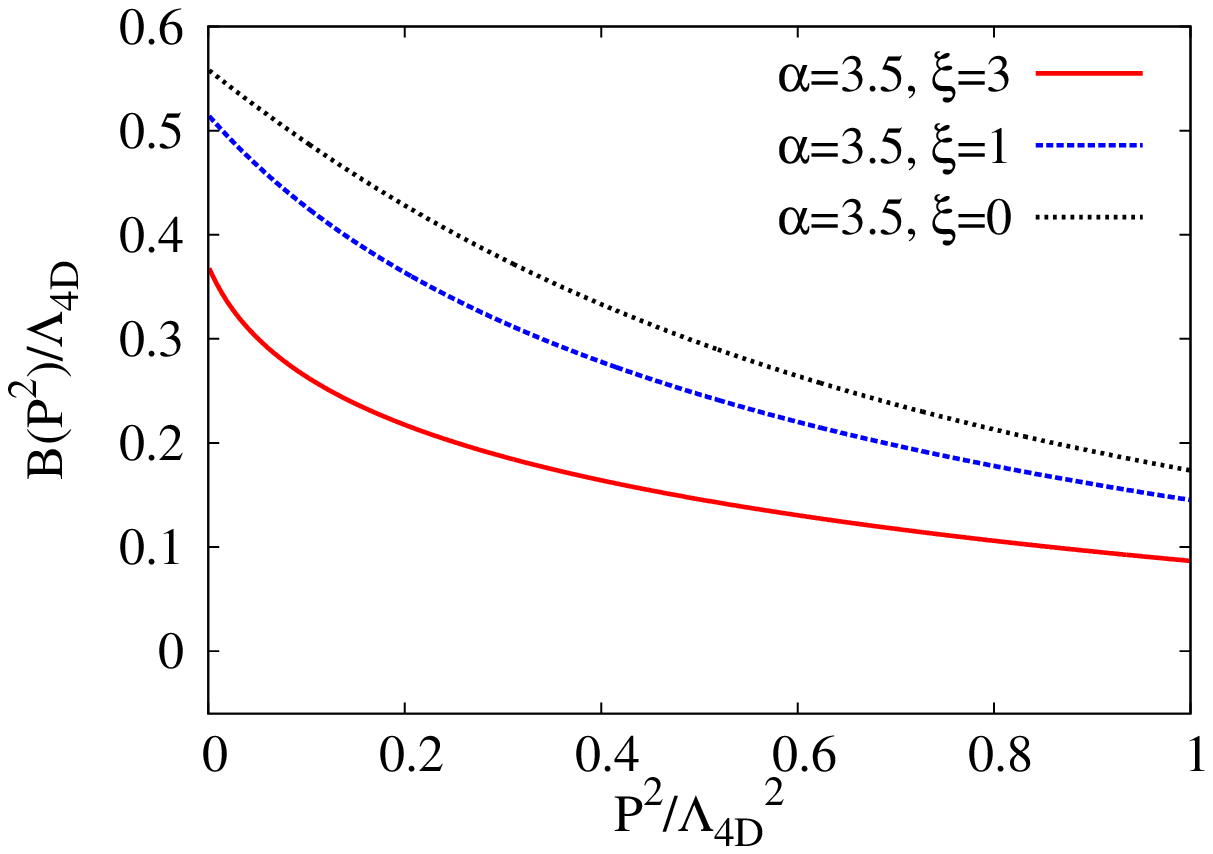}
  \caption{\label{fig:AB_25_35}
  Gauge dependence for $\alpha=2.5$ and $3.5$, with
  $m_0=0$, $\delta_{\rm 4D} = 0.01\Lambda_{\rm 4D}$.}
\end{center}
\end{figure}

We show the numerical results of $A(P^2)$ and $B(P^2)$
in Fig.~\ref{fig:AB_25_35}.
One sees that $A(p^2)$ increases with respect to $\xi$.  This can easily
be understood because $A$ has the form of $A=1+{\mathcal O}(\xi)$, so it
becomes larger when $\xi$ increases. On the other hand, $B(p^2)$
decreases when $\xi$ becomes larger. This can also be understood by
following discussion; $B$ has the form of $B \propto (3+\xi) \int \md
Q F(Q) \Delta(Q)$, and although $3+\xi$ becomes larger with increasing
$\xi$, $\Delta(Q) = 1/(A^2 Q^2 + B^2)$ decreases when $A$ increases.
Consequently, $B$ has smaller value for larger $\xi$. Note that $A=1$
always persists in the case $\xi=0$ as obviously read from
Eq.~(\ref{eq:original_A}), which is the well-known consequence of
choosing the Landau gauge. We also studied different
values of $\alpha$, and found that the above mentioned tendency did
not change, so we only showed the results with $\alpha=2.5$ and $3.5$
here.

\begin{figure}[!h]
\begin{center}
  \includegraphics[width=6.7cm,keepaspectratio]{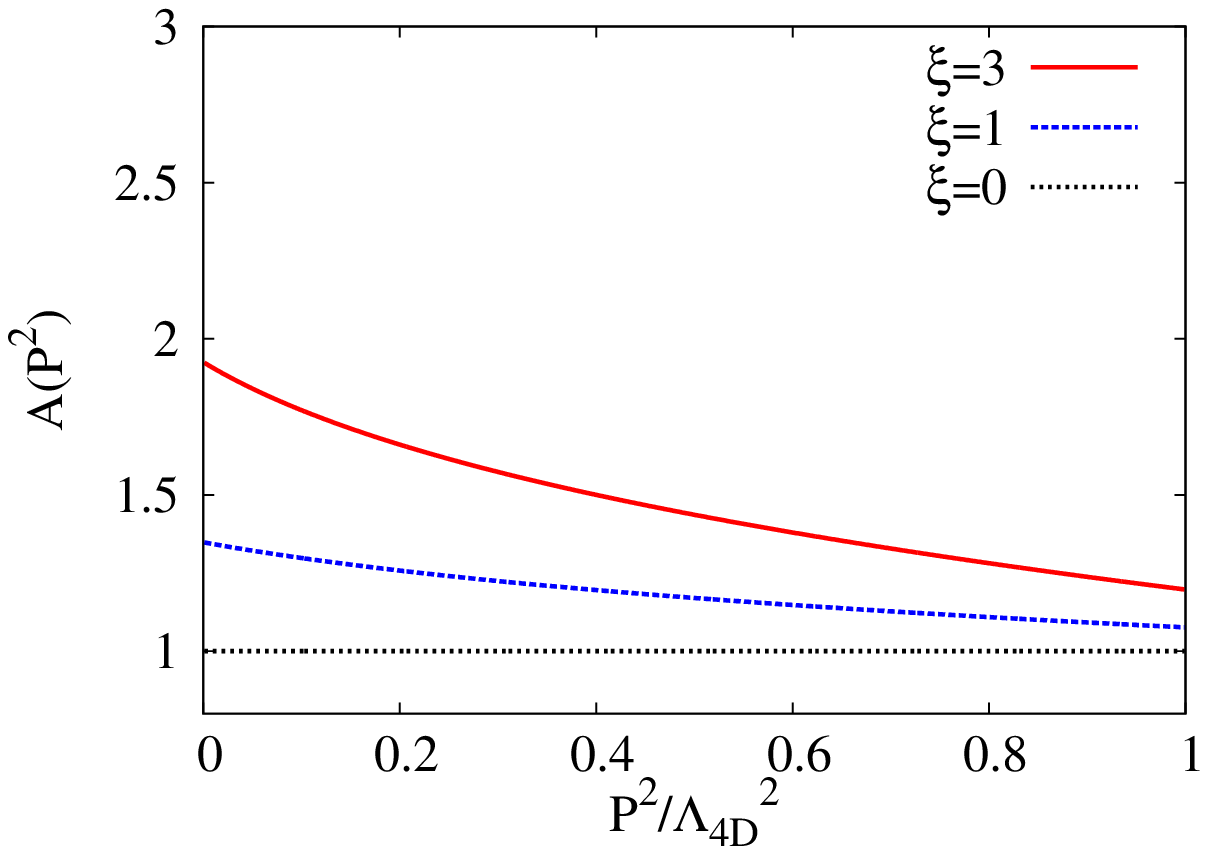}
  \includegraphics[width=6.7cm,keepaspectratio]{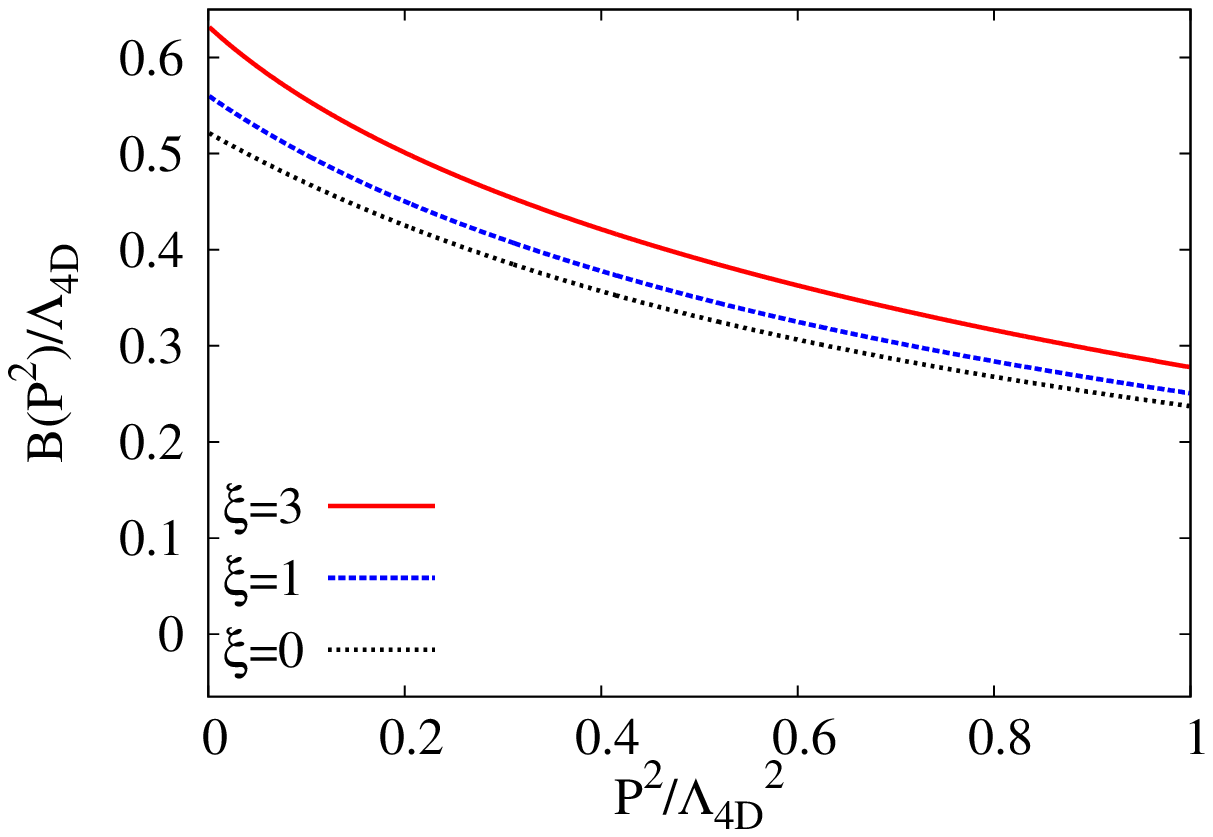}
  \caption{\label{fig:ABm_25}
  Results for $\alpha=2.5$, with
  $m_0=0.1\Lambda_{\rm 4D}$,
  $\delta_{\rm 4D} = 0.01\Lambda_{\rm 4D}$.}
\end{center}
\end{figure}

It may also be worth studying the case with finite $m_0$.
Figure~\ref{fig:ABm_25} shows the numerical results
with $m_0 = 0.1 \Lambda_{\rm 4D}$.
One notes that the obtained values of $A$ and $B$ are closer for
various gauge comparing to the case with $m_0=0$. We can read
from the results that the value of $B$ is dominated by the factor
$3+\xi$ rather than $\Delta(Q)$ because $A$ has closer values
for this case as seen in the left panel of Fig~\ref{fig:ABm_25}.

\subsection{\label{subsec:re_3d}
Solutions with the 3D cutoff scheme}
Here we show the numerical results of $C(p_0,p)$, $A(p_0,p)$ and
$B(p_0,p)$ for various values of $\xi$ with the three dimensional cutoff.
\begin{figure}[!h]
\begin{center}
  \includegraphics[width=5.4cm,keepaspectratio]{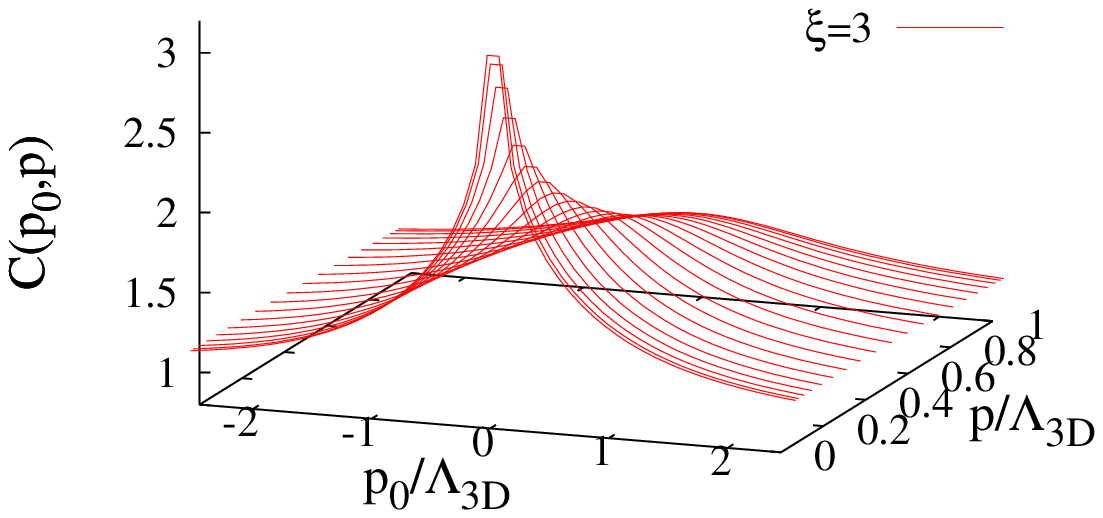}
  \includegraphics[width=5.4cm,keepaspectratio]{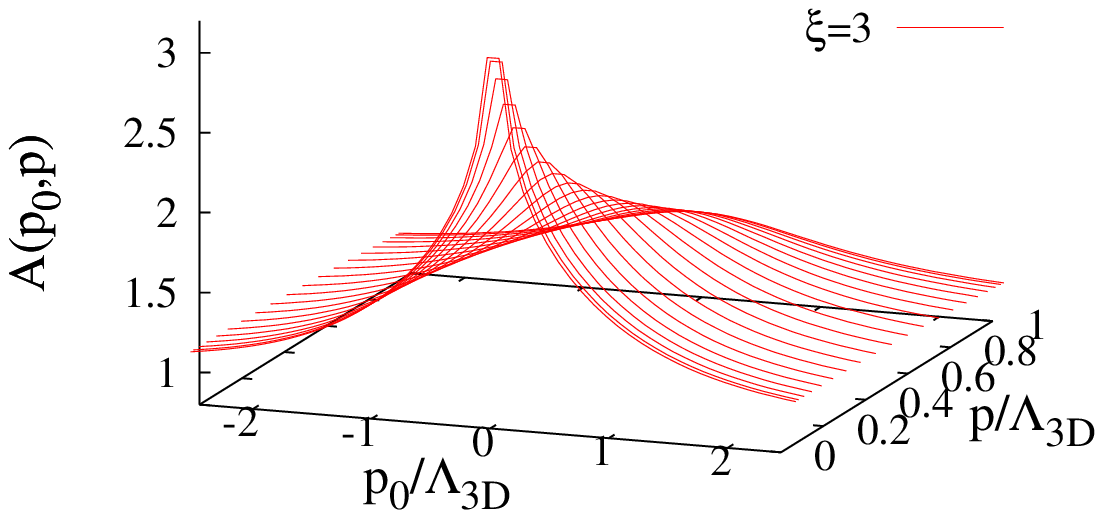}
  \includegraphics[width=5.4cm,keepaspectratio]{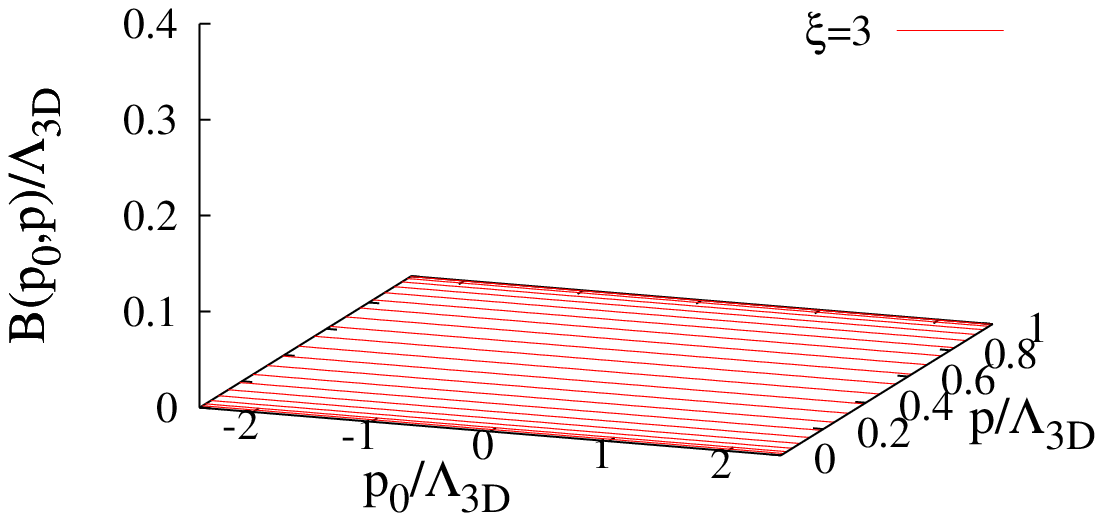}
  \includegraphics[width=5.4cm,keepaspectratio]{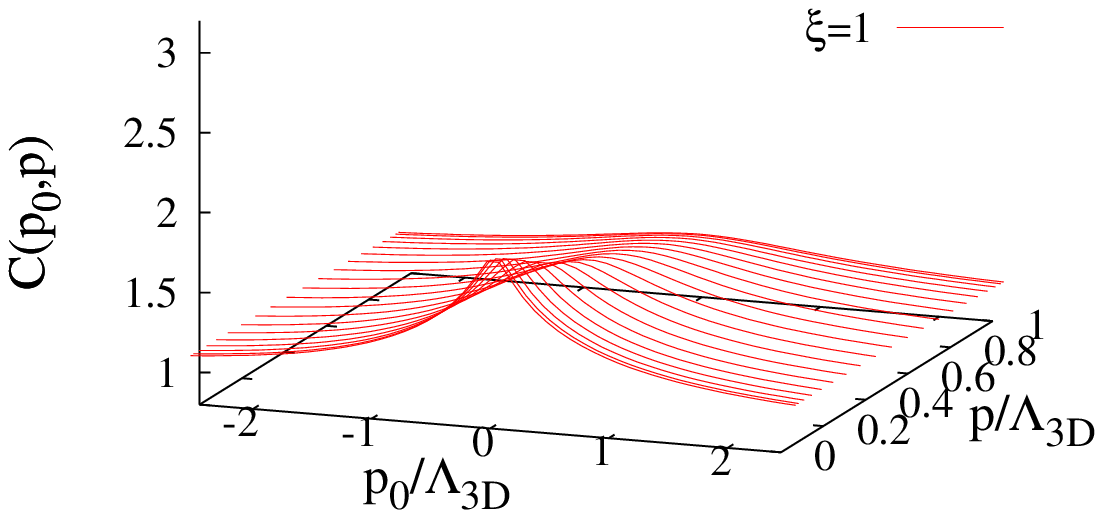}
  \includegraphics[width=5.4cm,keepaspectratio]{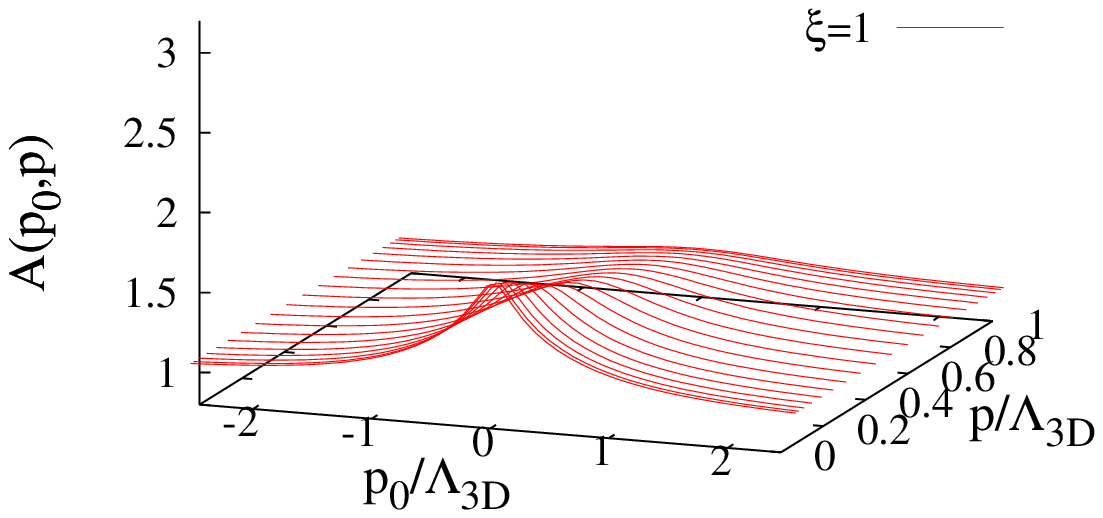}
  \includegraphics[width=5.4cm,keepaspectratio]{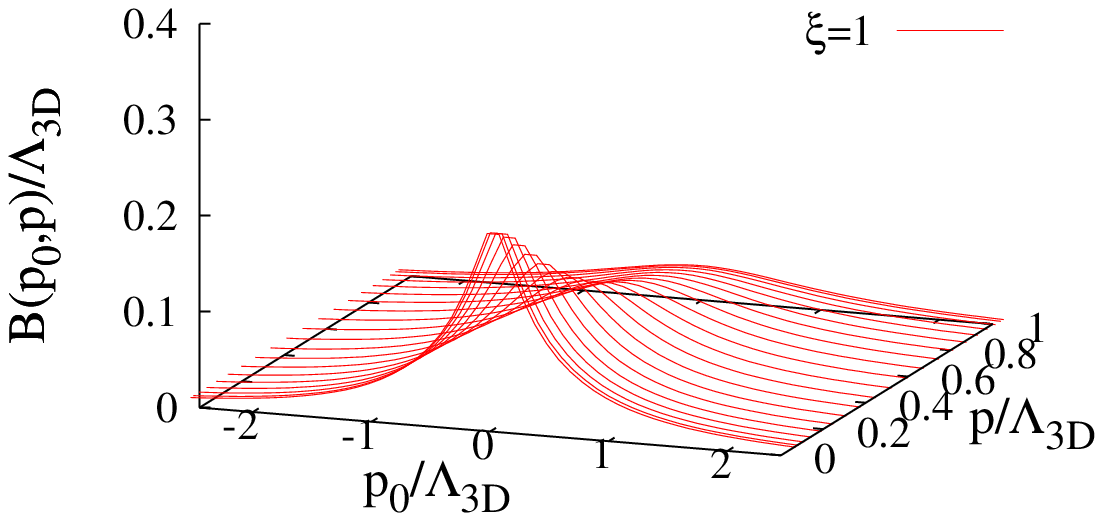}
  \includegraphics[width=5.4cm,keepaspectratio]{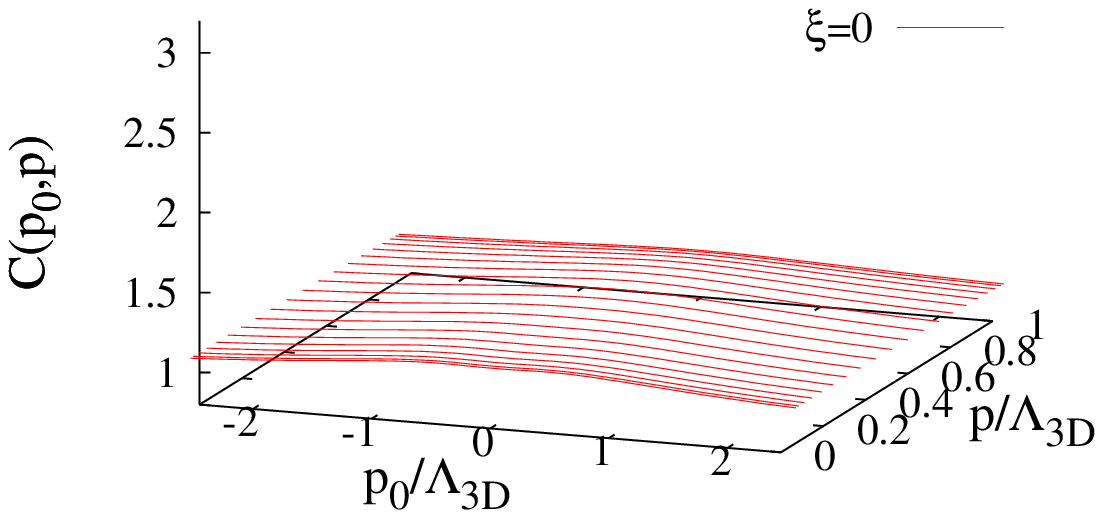}
  \includegraphics[width=5.4cm,keepaspectratio]{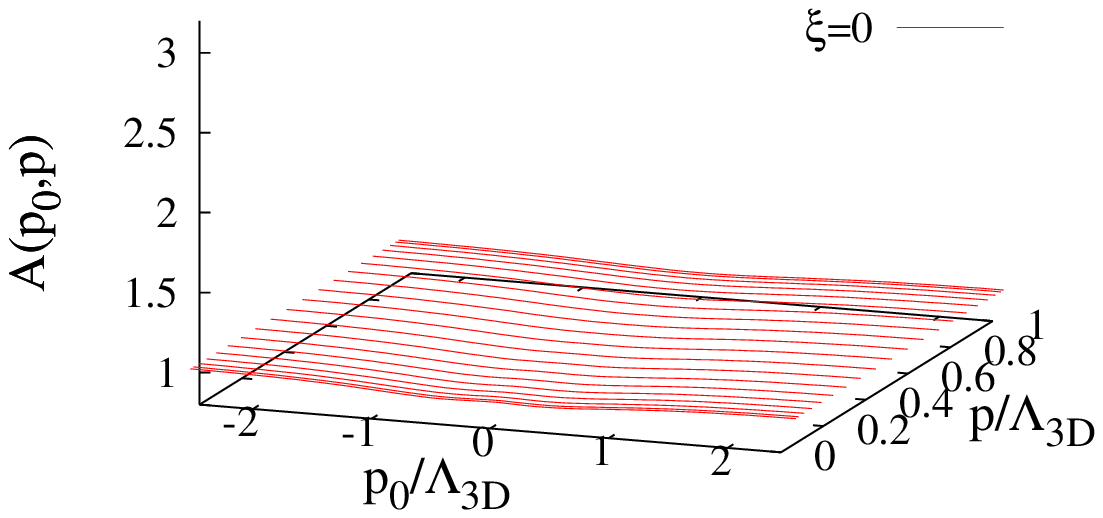}
  \includegraphics[width=5.4cm,keepaspectratio]{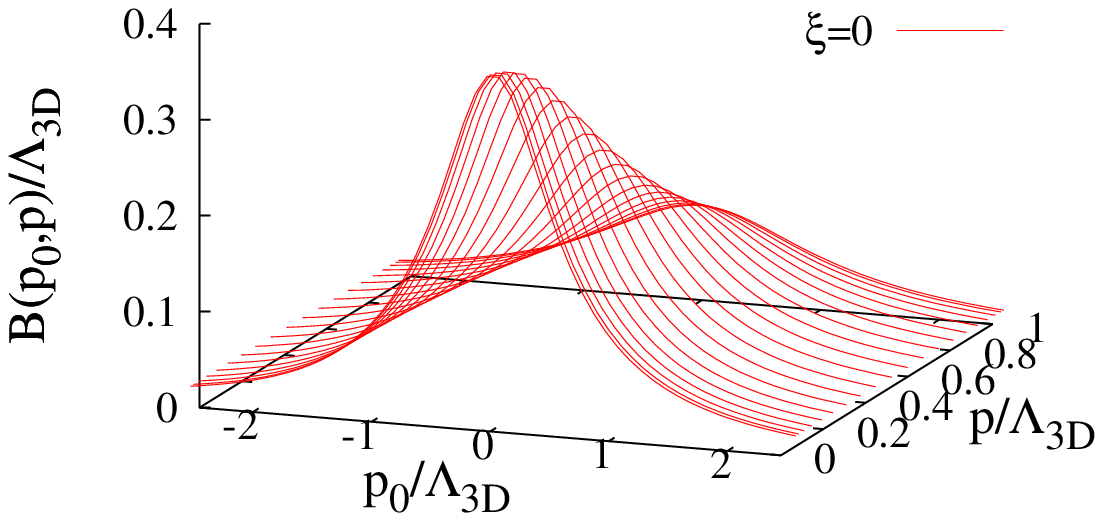}
  \caption{\label{fig:ABC_25}
  Gauge dependence for $\alpha=2.5$, with $m_0=0$ and
  $\delta_3 = 0.01\Lambda_{\rm 3D}$.}
\end{center}
\end{figure}
\begin{figure}[!h]
\begin{center}
  \includegraphics[width=5.4cm,keepaspectratio]{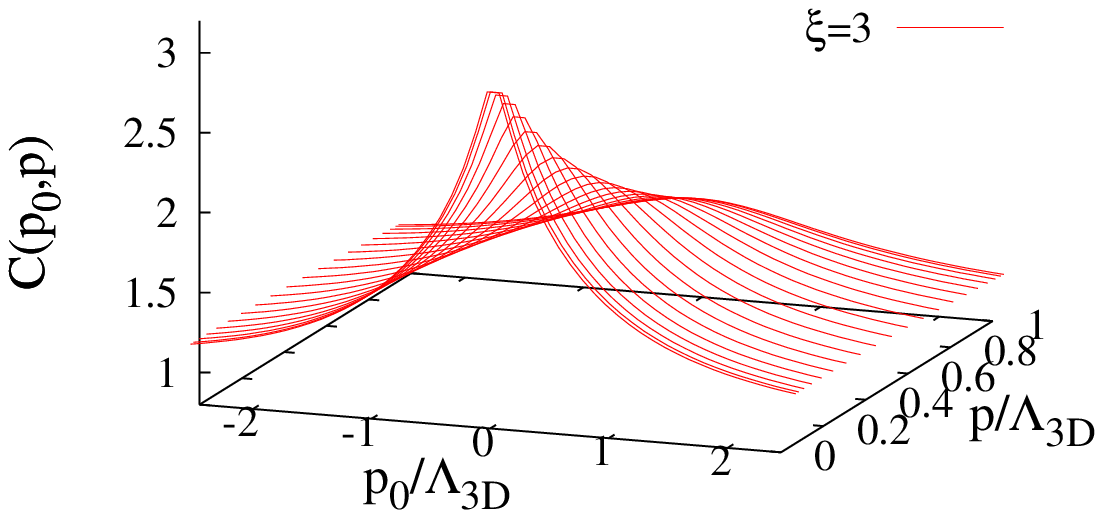}
  \includegraphics[width=5.4cm,keepaspectratio]{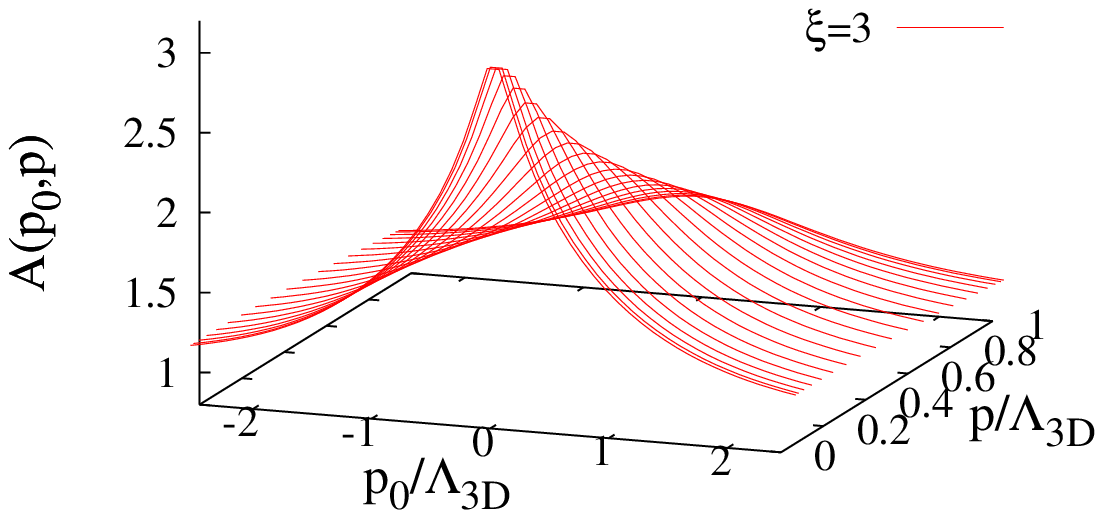}
  \includegraphics[width=5.4cm,keepaspectratio]{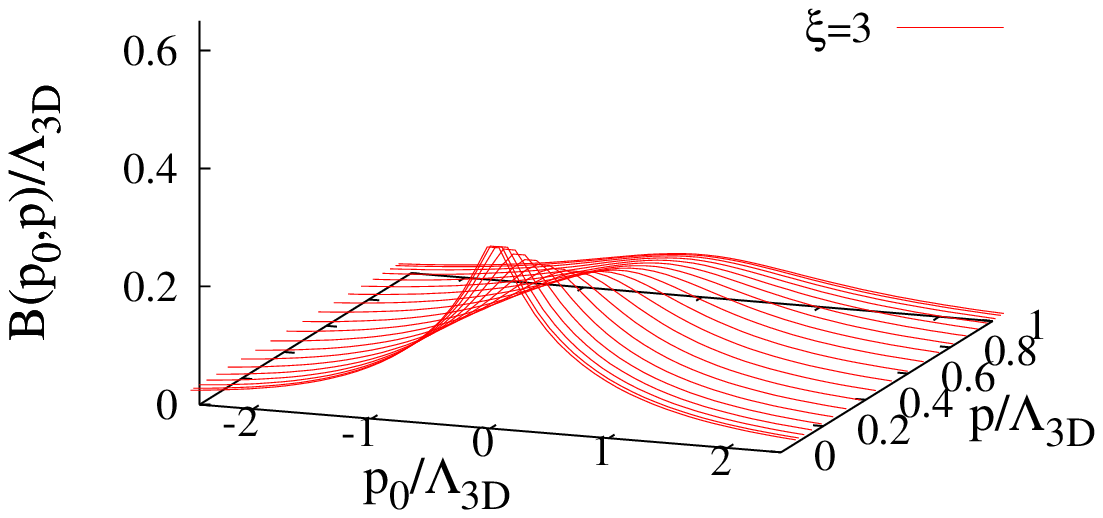}
  \includegraphics[width=5.4cm,keepaspectratio]{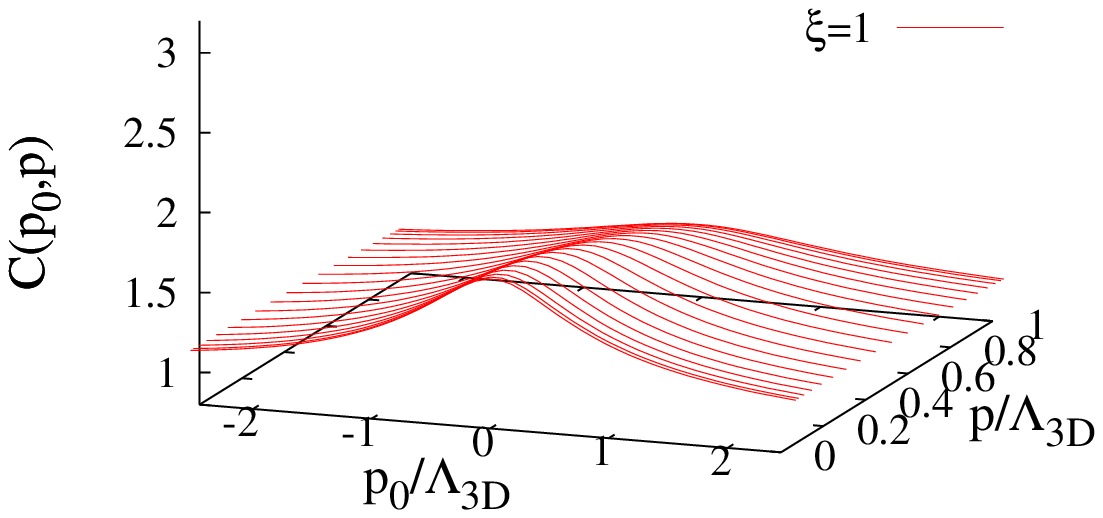}
  \includegraphics[width=5.4cm,keepaspectratio]{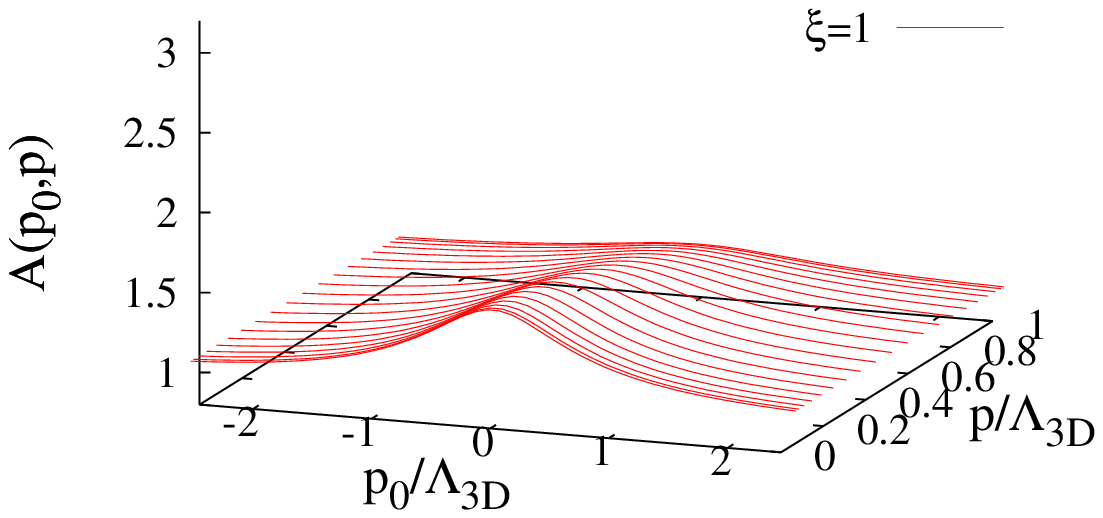}
  \includegraphics[width=5.4cm,keepaspectratio]{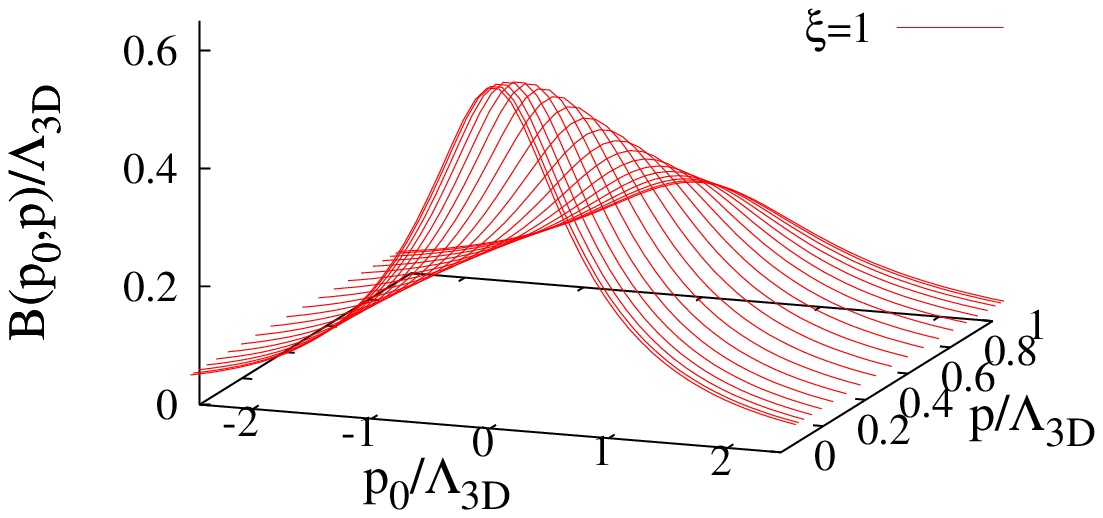}
  \includegraphics[width=5.4cm,keepaspectratio]{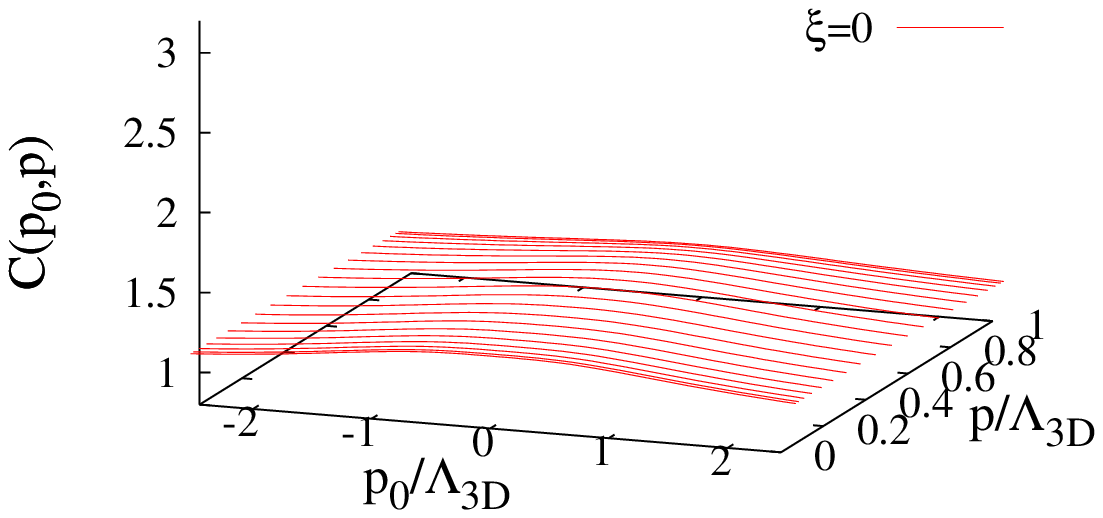}
  \includegraphics[width=5.4cm,keepaspectratio]{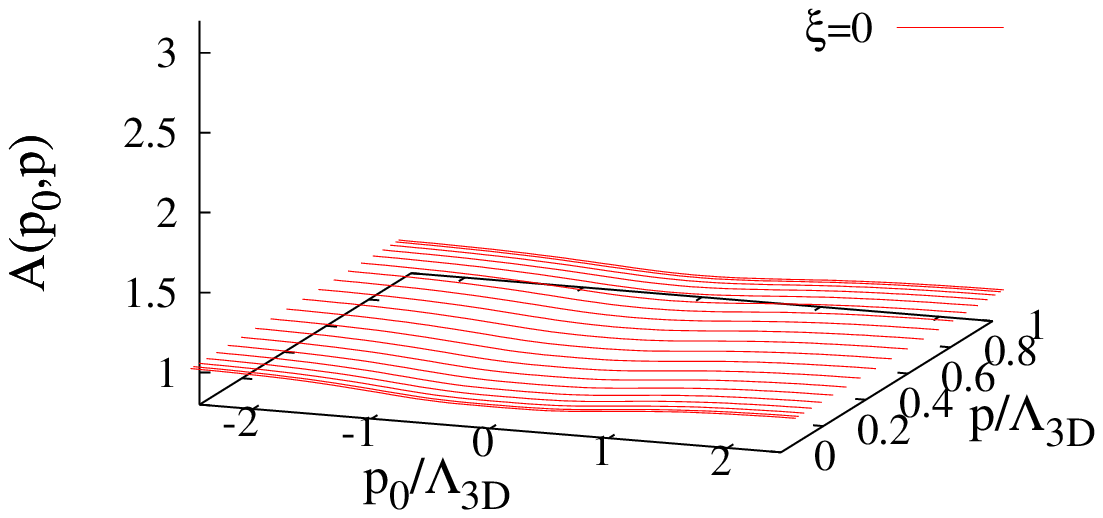}
  \includegraphics[width=5.4cm,keepaspectratio]{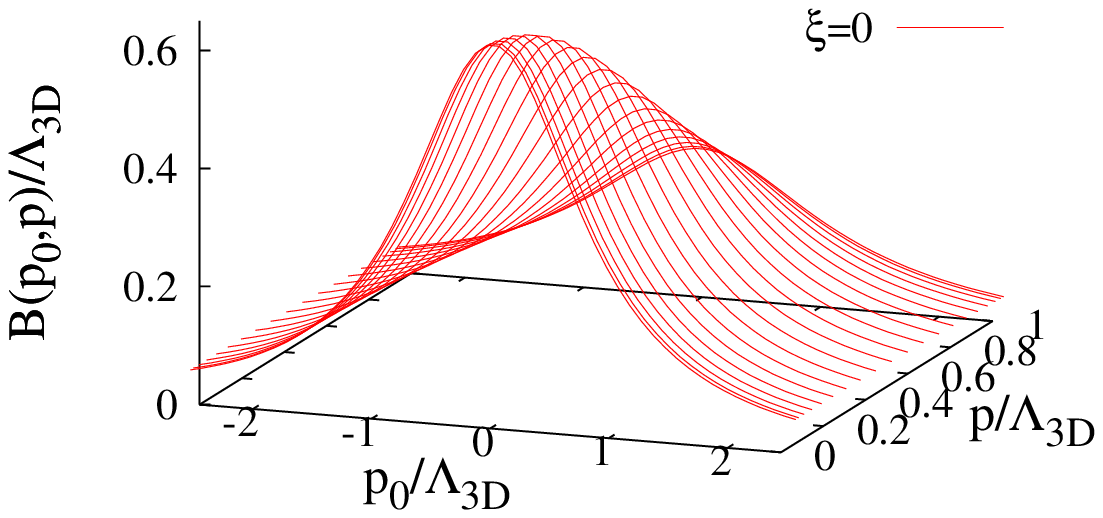}
  \caption{\label{fig:ABC_35}
  Gauge dependence for $\alpha=3.5$, with $m_0=0$ and
  $\delta_3 = 0.01\Lambda_{\rm 3D}$.}
\end{center}
\end{figure}
Figures~\ref{fig:ABC_25} and \ref{fig:ABC_35} display the solutions
for $\alpha=2.5$ and $3.5$ with the gauges, $\xi=0$, $1$ and
$3$. In performing the integral in $q_0$ direction, we set the lower and
upper limit as $\int_{-\Lambda_0}^{\Lambda_0} \md q_0$ with 
$\Lambda_0 = 10\Lambda_{\rm 3D}$ because it is technically difficult
to directly take the infinite range. We have numerically confirmed that
the solutions are not affected by the choice of the cutoff $\Lambda_0$ if
we take large enough value for it such as
$\Lambda_0 > 5\Lambda_{\rm 3D}$.

From the obtained curves, one sees the similar tendencies that the
renormalization factors $A$ and $C$ are large when $\xi$ is large,
while the mass factor $B$ is small for larger $\xi$. It may be interesting
to note that $A$ and $C$ are close to $1$, but deviate from $1$
in the case of the Landau gauge, $\xi =0$, which comes from the
effect of the separation of the 4D momentum $P$ to 3D momentum
$(p_0,p)$.

\begin{figure}[!h]
\begin{center}
  \includegraphics[width=5.4cm,keepaspectratio]{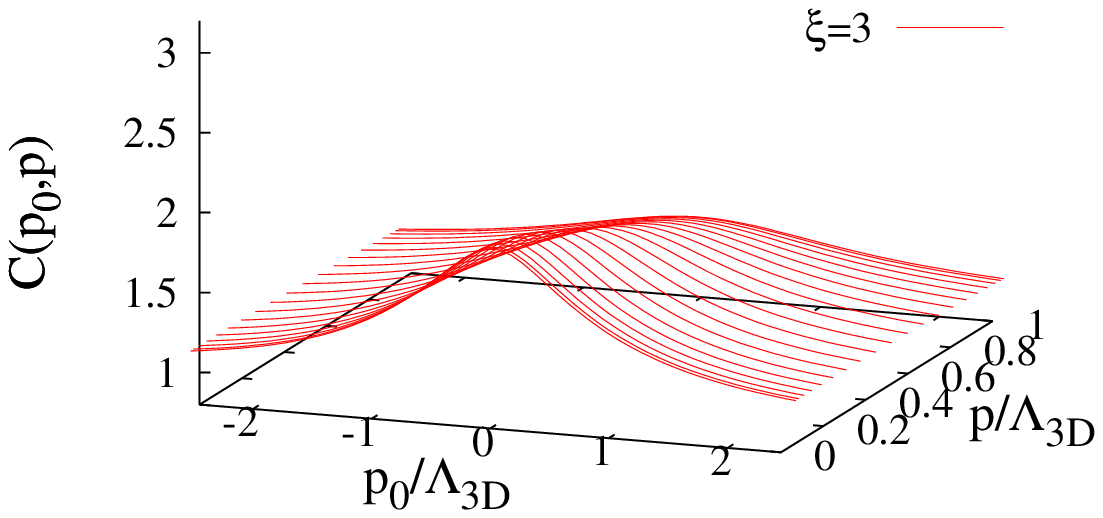}
  \includegraphics[width=5.4cm,keepaspectratio]{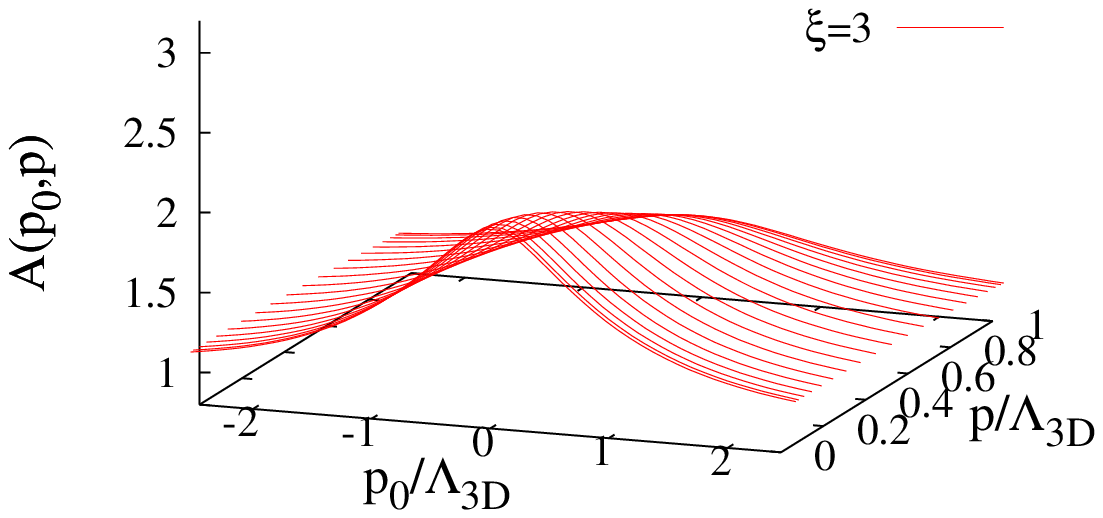}
  \includegraphics[width=5.4cm,keepaspectratio]{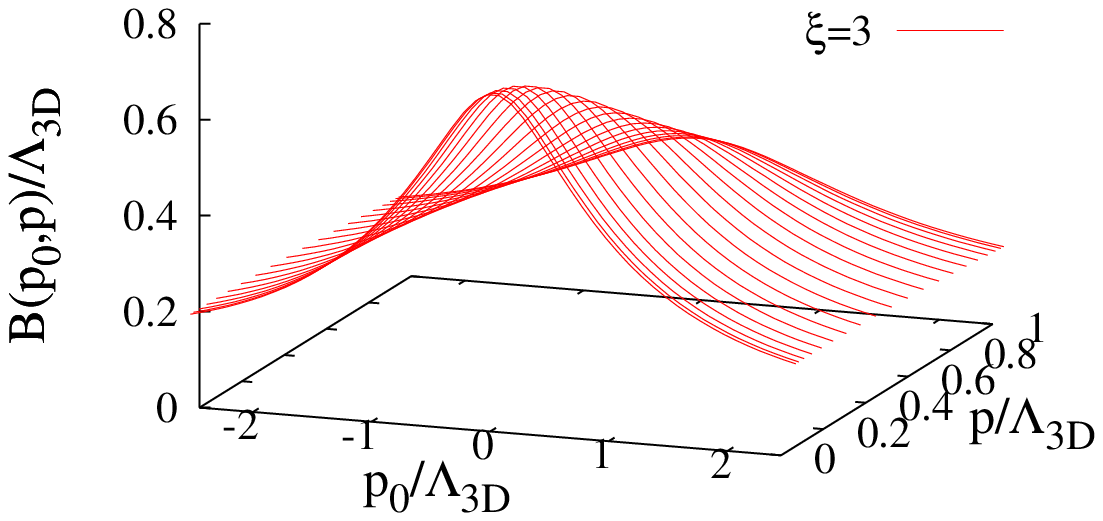}
  \includegraphics[width=5.4cm,keepaspectratio]{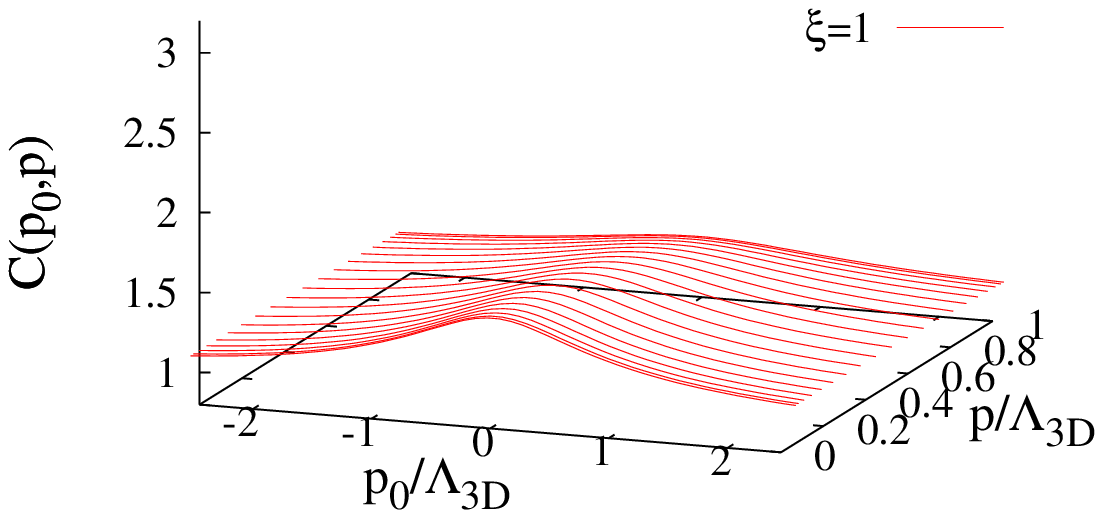}
  \includegraphics[width=5.4cm,keepaspectratio]{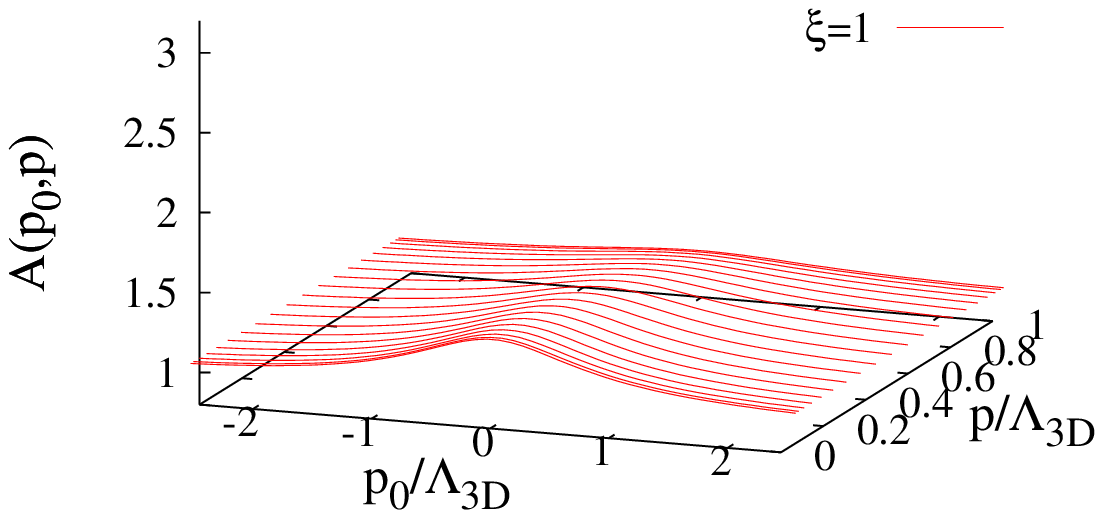}
  \includegraphics[width=5.4cm,keepaspectratio]{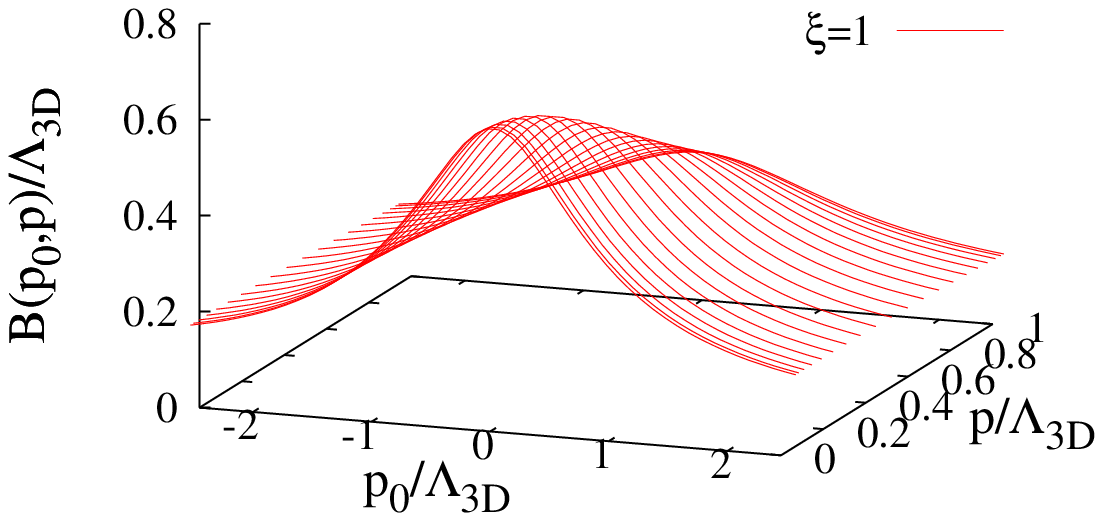}
  \includegraphics[width=5.4cm,keepaspectratio]{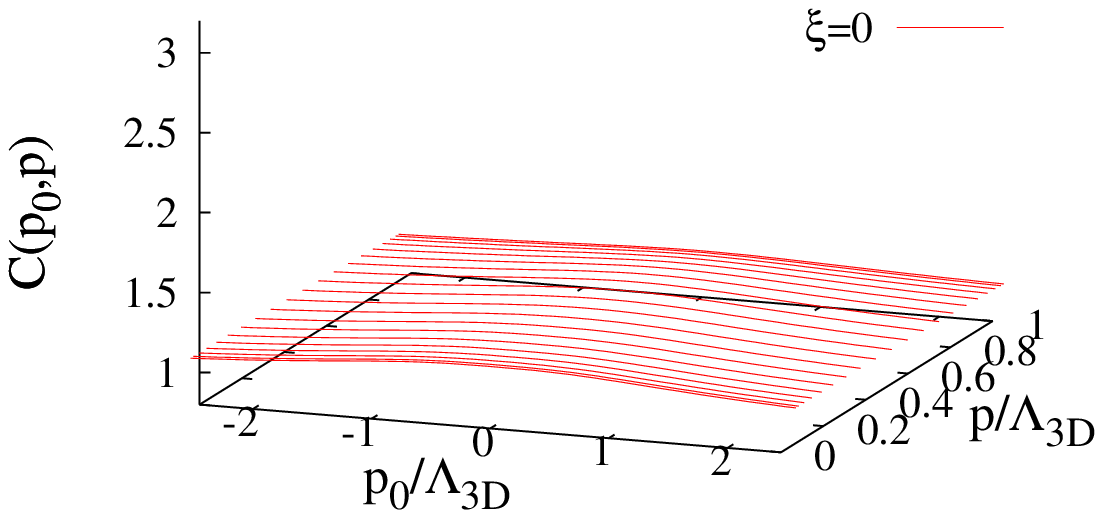}
  \includegraphics[width=5.4cm,keepaspectratio]{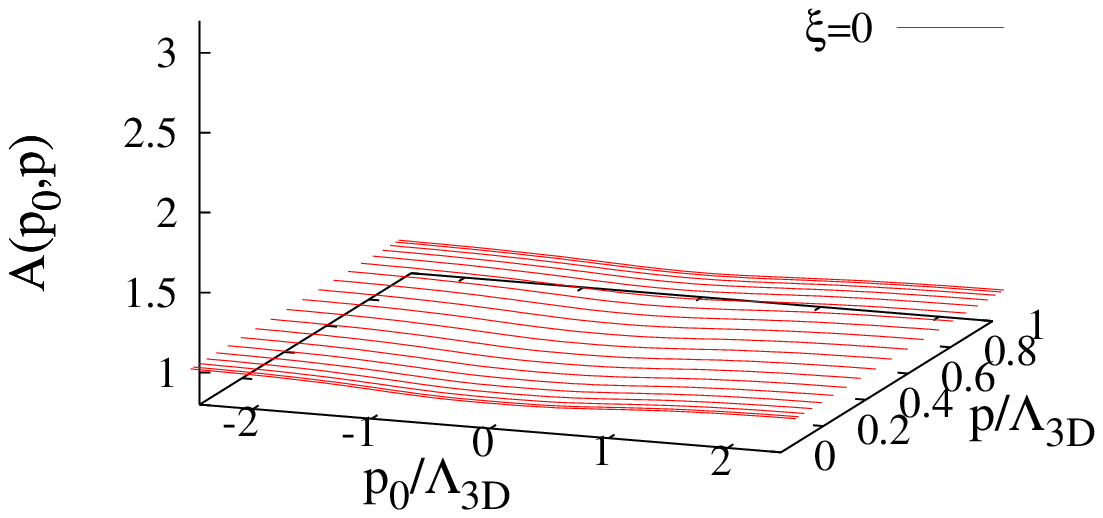}
  \includegraphics[width=5.4cm,keepaspectratio]{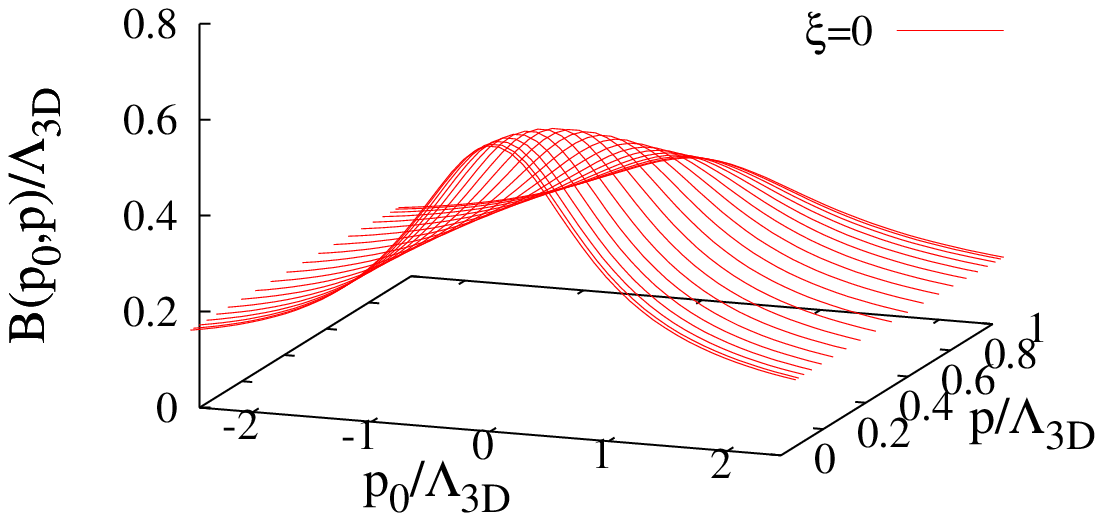}
  \caption{\label{fig:ABC_25m}
  Results for $\alpha=2.5$, with $m_0=0.1\Lambda_{\rm 3D}$ and
  $\delta_3 = 0.01\Lambda_{\rm 3D}$.}
\end{center}
\end{figure}
We next check the effect of the finite bare mass $m_0 \neq 0$.
Figure \ref{fig:ABC_25m} shows the solutions with
$m_0=0.1\Lambda_{\rm 3D}$ and $\alpha=2.5$ for various
gauges. Again, one sees the similar qualitative
tendency with the 4D case; the renormalization factors $A$ and $C$
become close to $1$ comparing to the results with $m_0=0$, and
the mass factor $B$ increases with $\xi$.

\subsection{\label{subsec:comparison}
Comparison between 4D and 3D cutoff regularizations}
We have seen above that the qualitative feature of the solutions with
the 4D and 3D cutoff regularizations are similar. It may also be
interesting to show the quantitative comparison between these
regularizations, although the direct comparison is not possible since
the variables are different in two methods.
\begin{figure}[!h]
\begin{center}
  \includegraphics[width=6.7cm,keepaspectratio]{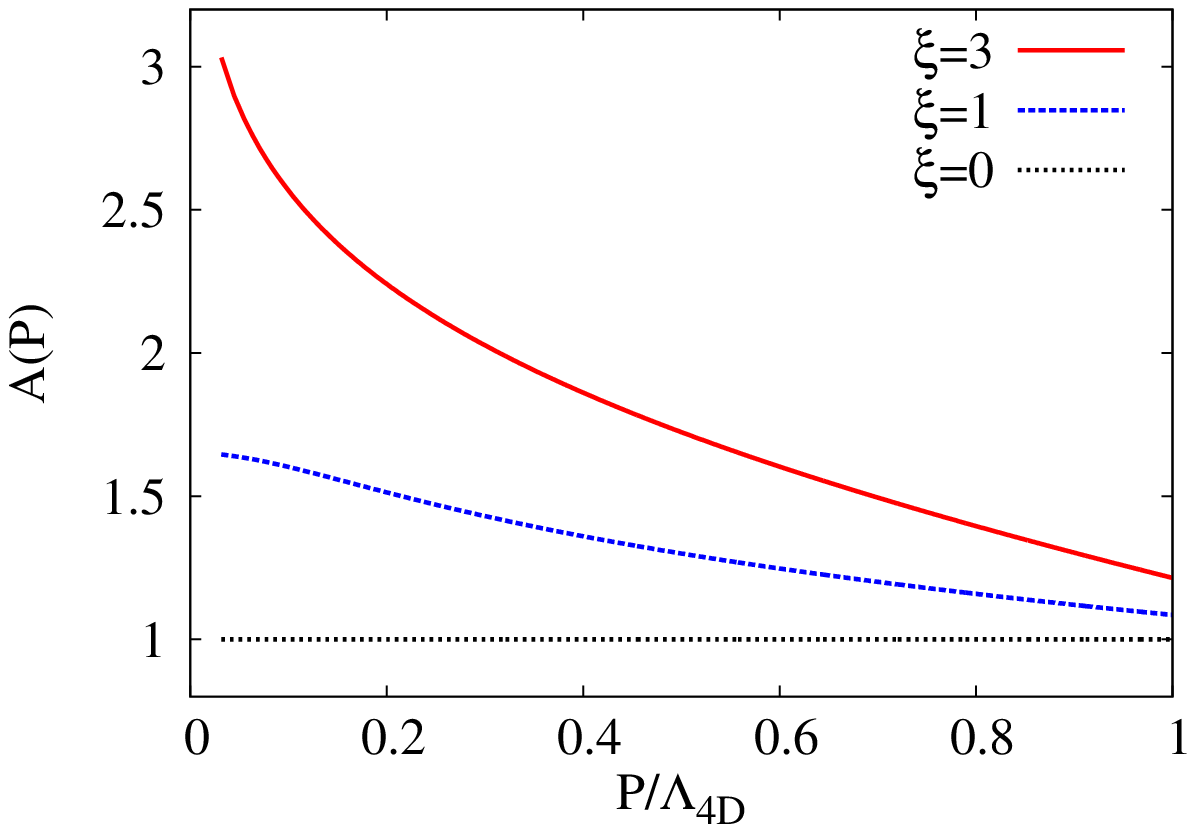}
  \includegraphics[width=6.7cm,keepaspectratio]{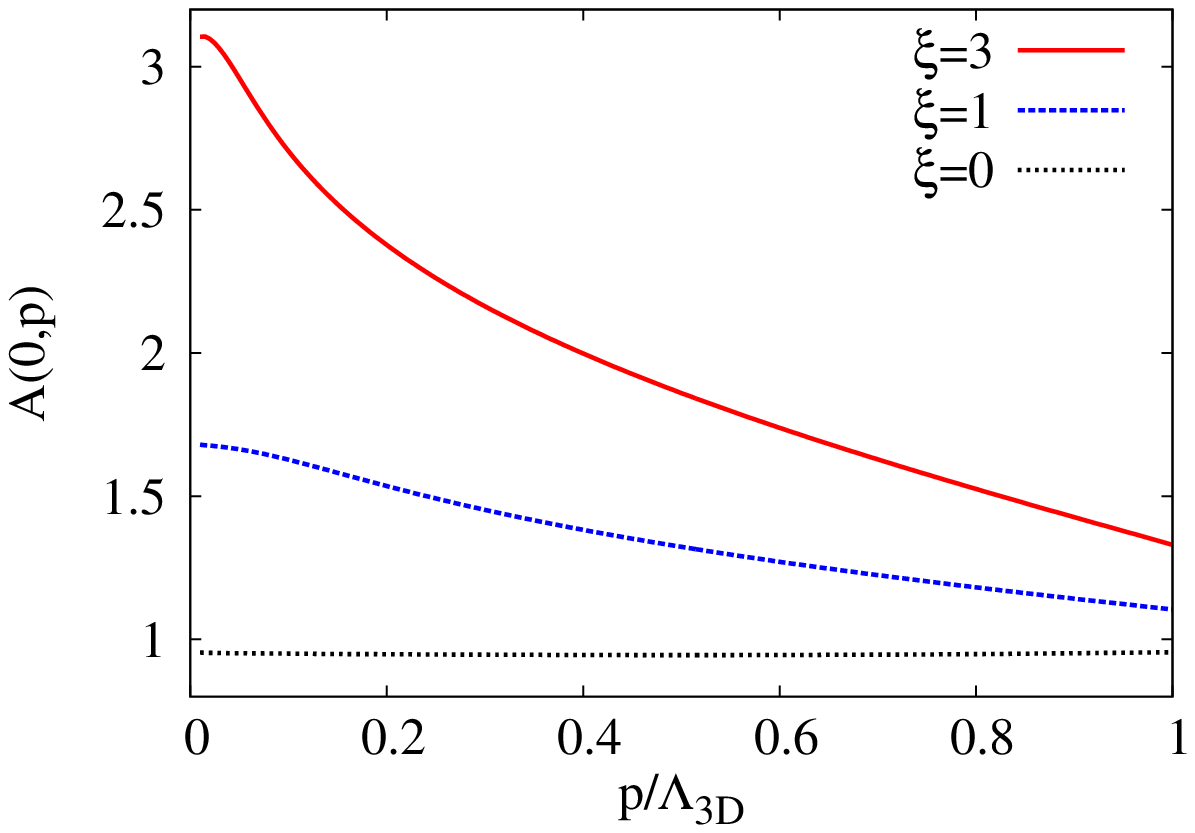}
  \includegraphics[width=6.7cm,keepaspectratio]{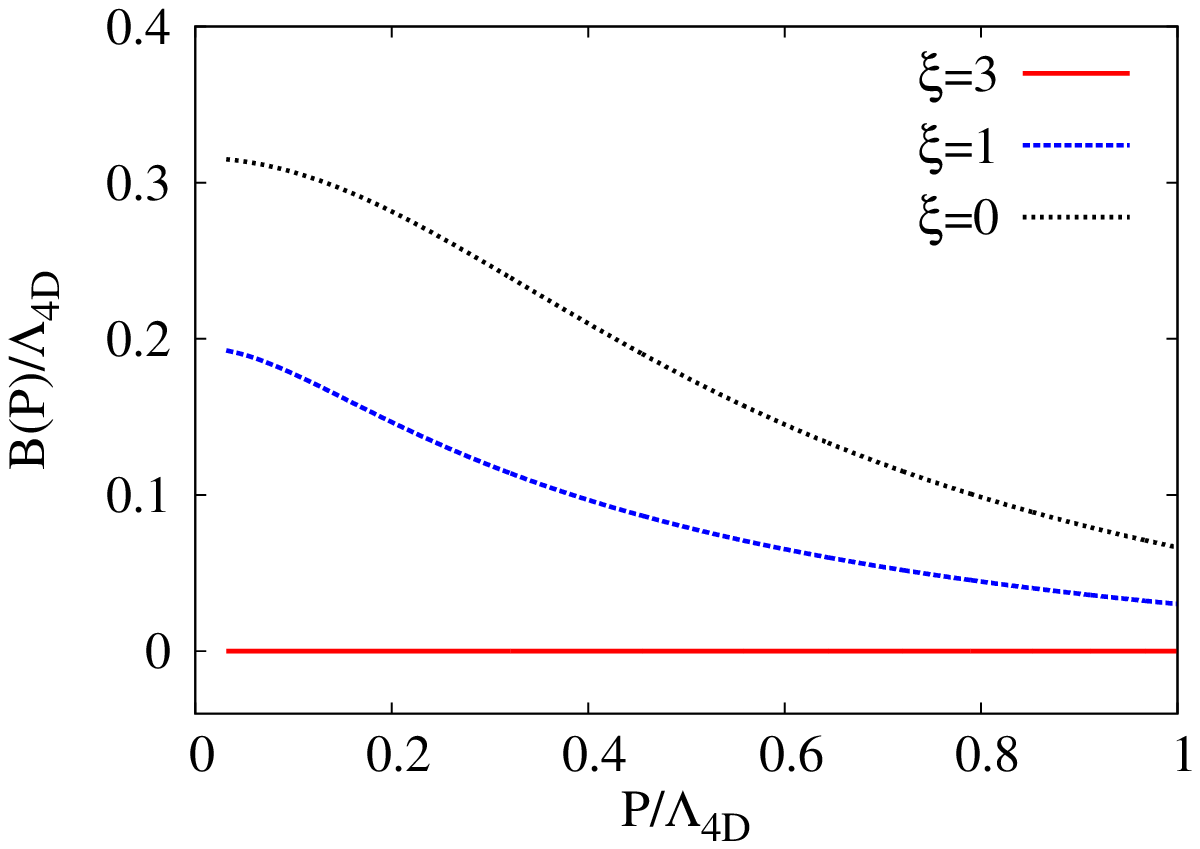}
  \includegraphics[width=6.7cm,keepaspectratio]{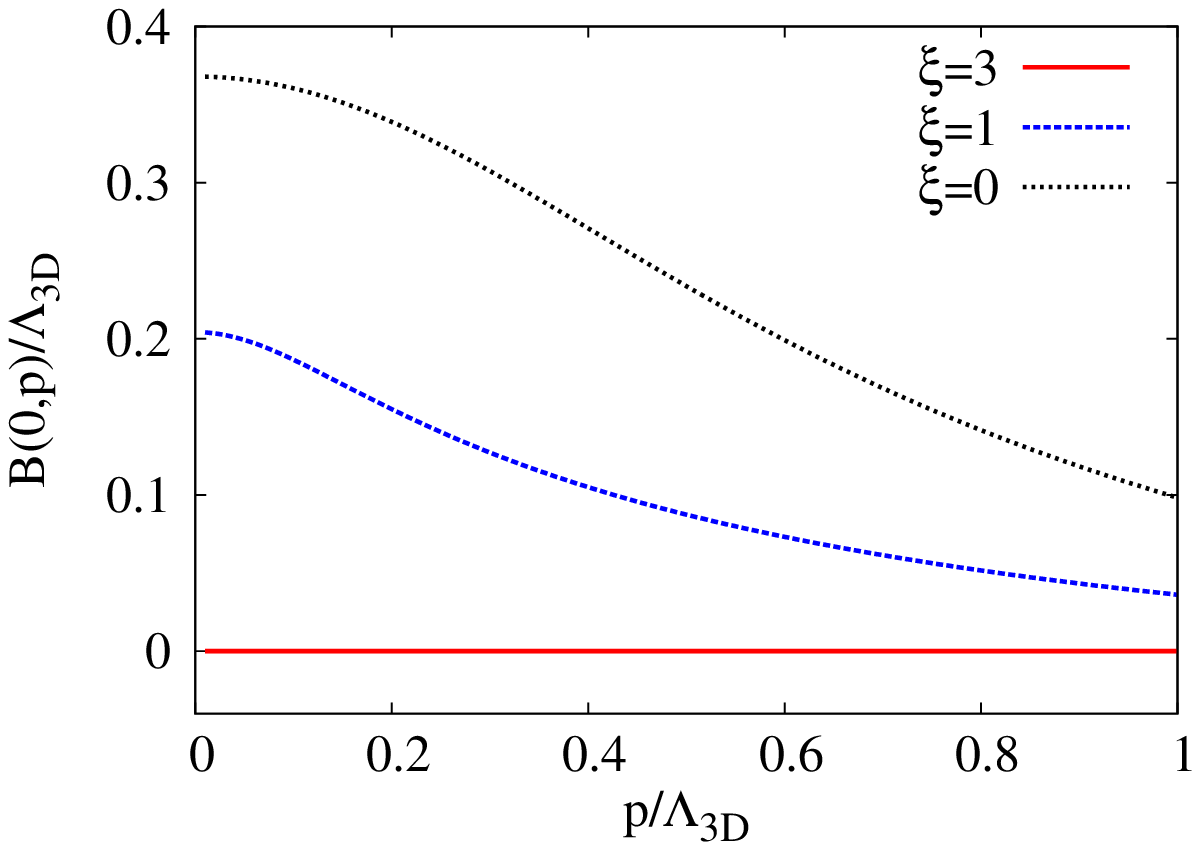}
  \caption{\label{fig:ABcomp_25}
  $A(P)$, $B(P)$ in the 4D cut (left), and
  $A(0,p)$, $B(0,p)$
  in the 3D cut (right) for $\alpha=2.5$.}
\end{center}
\end{figure}
\begin{figure}[!h]
\begin{center}
  \includegraphics[width=6.7cm,keepaspectratio]{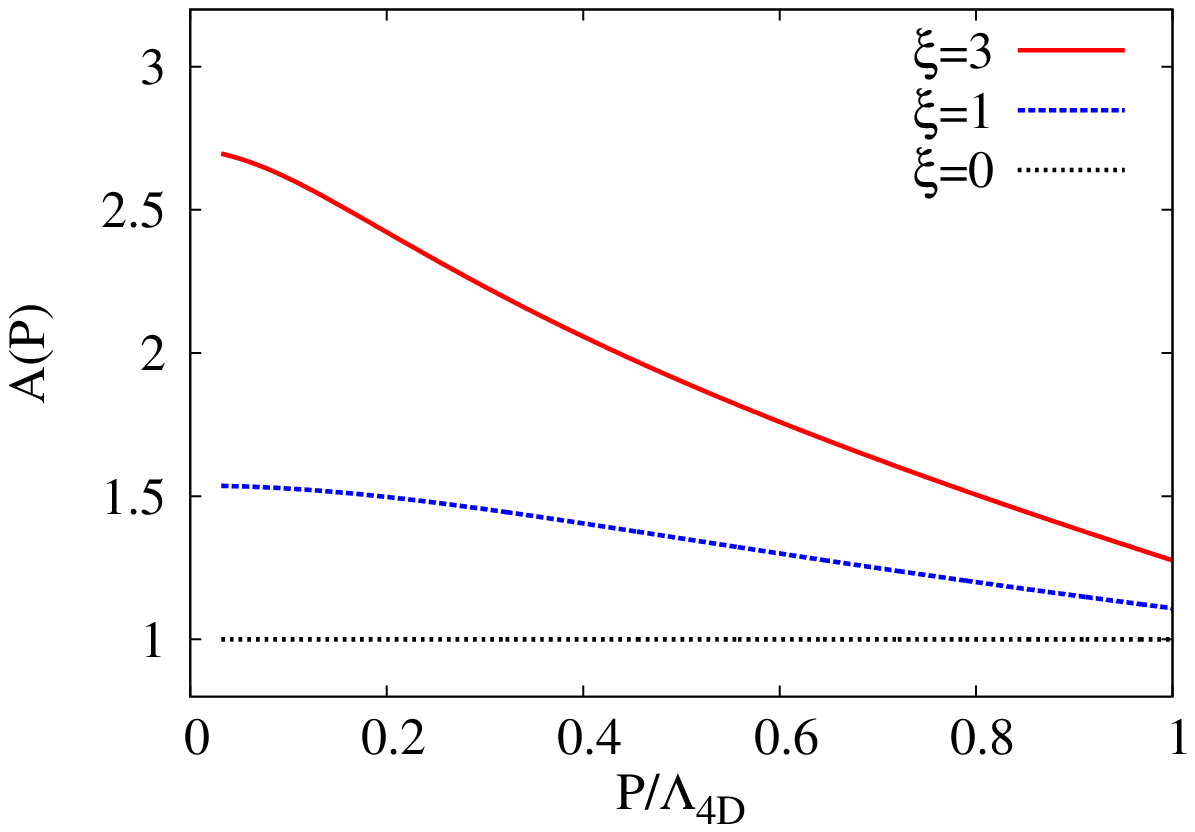}
  \includegraphics[width=6.7cm,keepaspectratio]{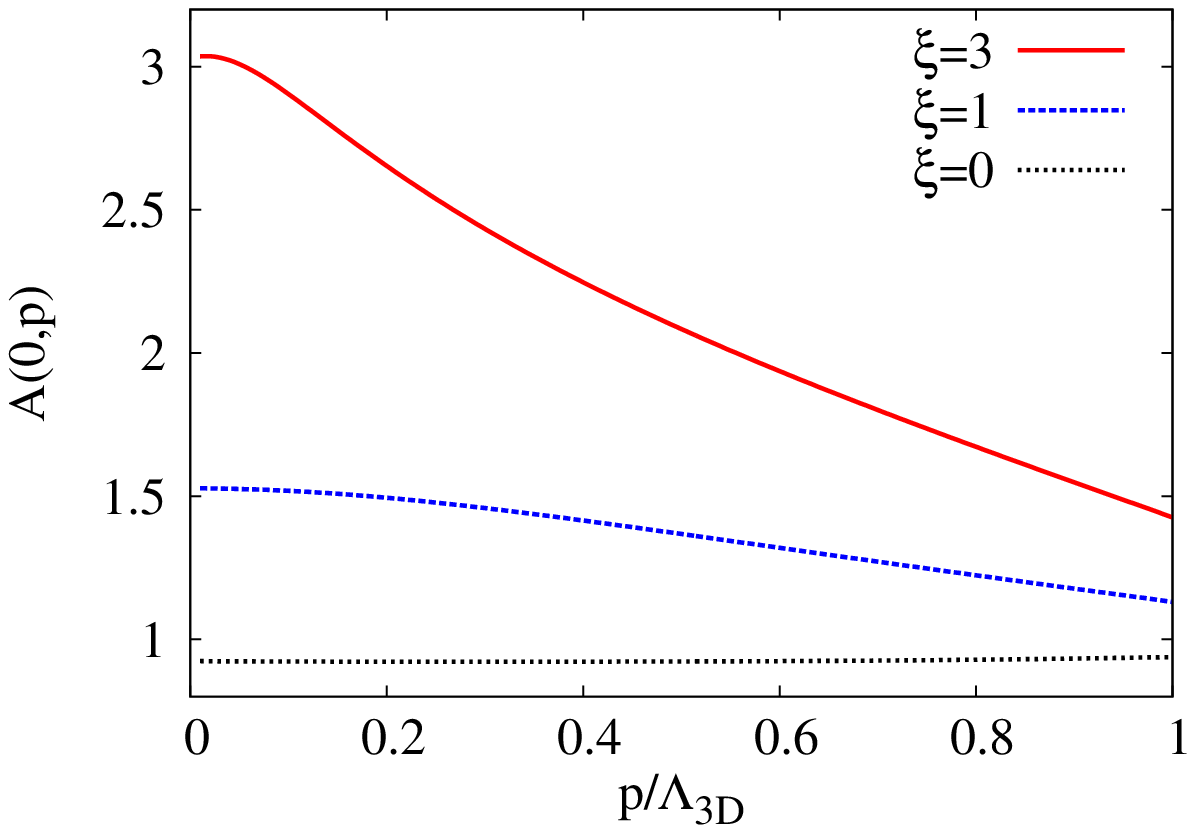}
  \includegraphics[width=6.7cm,keepaspectratio]{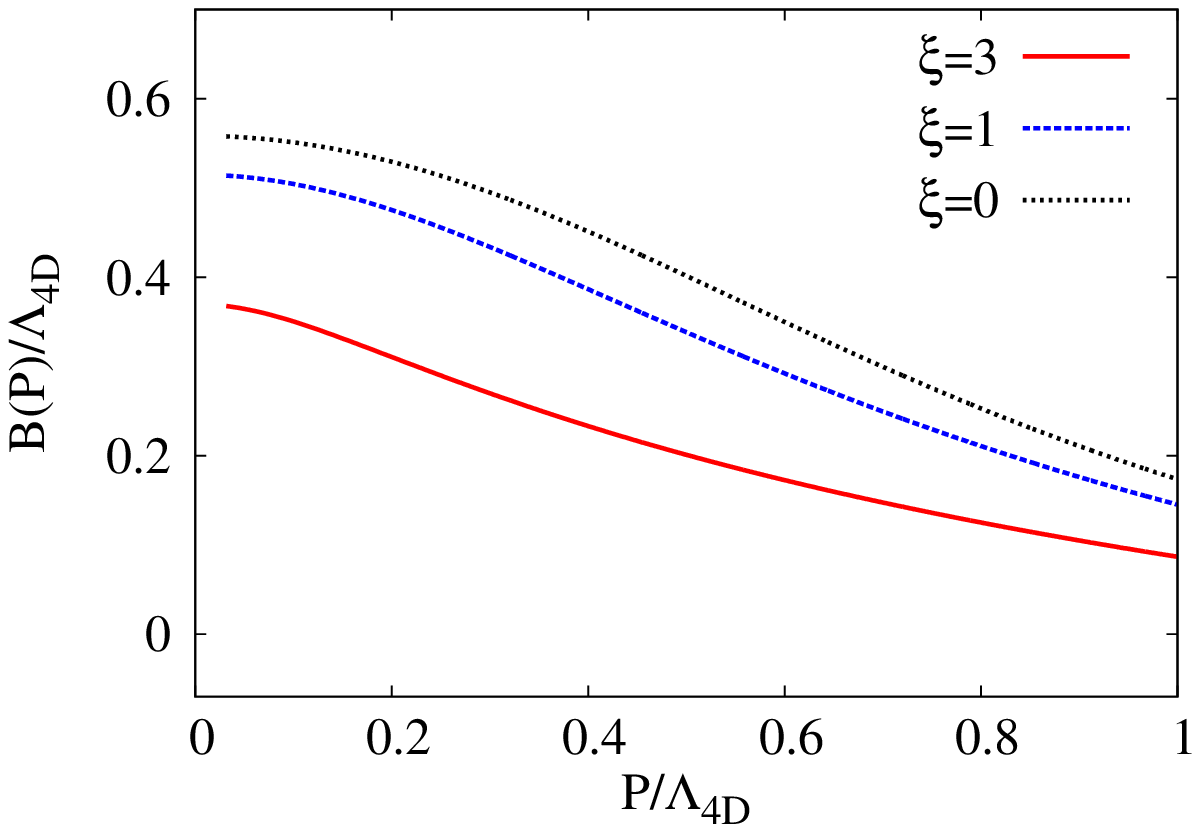}
  \includegraphics[width=6.7cm,keepaspectratio]{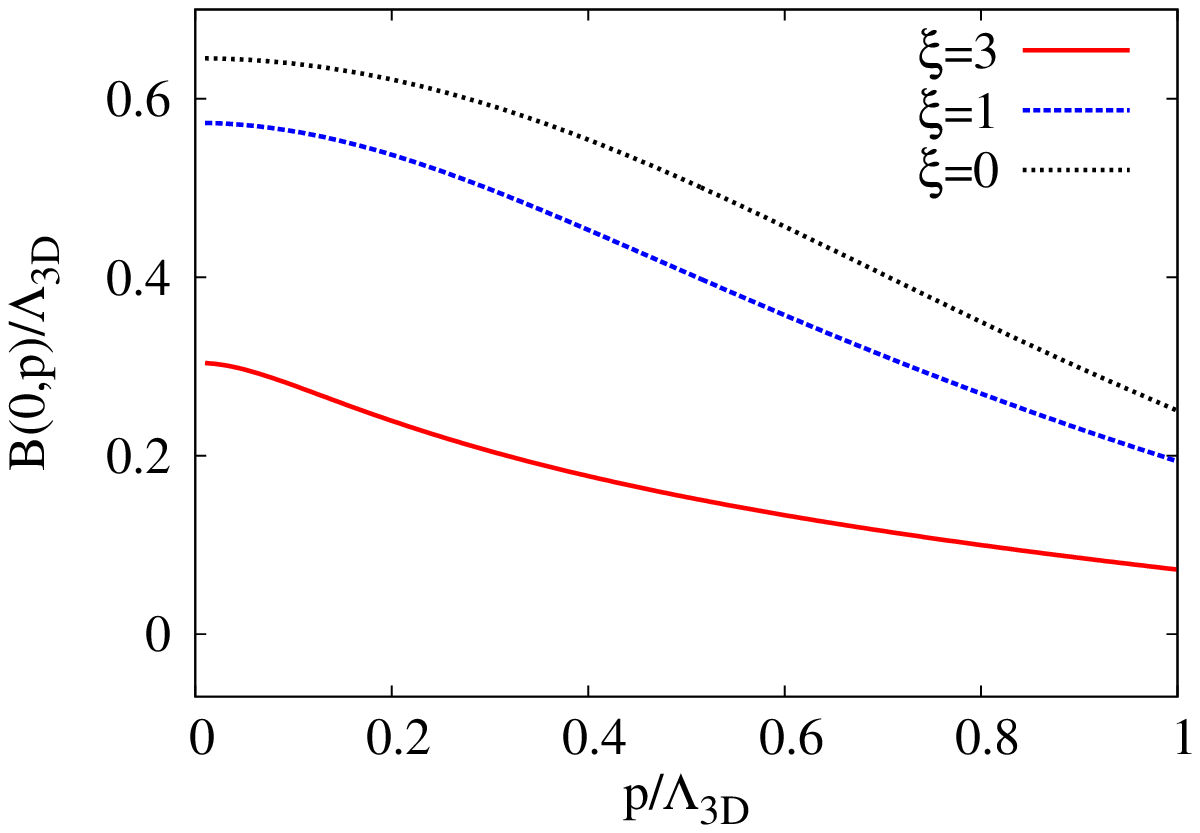}
  \caption{\label{fig:ABcomp_35}
  $A(P)$, $B(P)$ in the 4D cut (left), and
  $A(0,p)$, $B(0,p)$
  in the 3D cut (right) for $\alpha=3.5$.}
\end{center}
\end{figure}
In Figs. \ref{fig:ABcomp_25} and \ref{fig:ABcomp_35}, we align
the results of $A(P)$ and $B(P)$ as the functions of $P$ in the 4D
cutoff, and $A(0,p)$ and $B(0,p)$ as the functions of $p$ in the 3D
cutoff.

We find that the results between two regularizations do not alter
considerably, while the deviations between different gauges are
rather large. Then we read that, concerning on the solutions on
the momentum direction, the regularization dependence
is not serious compare to the gauge dependence.

\section{\label{sec:conclusion}
Concluding remarks}
We have studied the regularization dependence on the quenched SDE
in general gauge through applying the four and three dimensional
cutoff methods in this paper. The characteristic technical
difference lies on the number of variables where only one momentum,
$P$, exists in the 4D cutoff, on the other hand there appears two
different variables, $p_0$ and $p$, which makes the numerical analysis
challenging. The  obtained results between the 4D and 3D cutoff
methods show that the regularization dependence on the solutions
is not drastic comparing to the one on the gauge parameter. This
indicates that both the regularization  prescriptions can nicely be
adopted for the analysis on the SDE.

The gauge dependence seen in the results are due to the applied
approximations in deriving the equations. Especially, the tree-level
approximated form of the photon propagator in Eq. (\ref{eq:photon_prop})
is crucial when we consider the gauge dependence since the gauge
parameter manifestly appears in the equations. Therefore, for the
sake of obtaining the gauge independent solutions as indicated by
gauge theories, the general analyses, such as the ones based on the
unquenched equations~\cite{Kizilersu:2014ela, Oliensis:1990sg},
and the generalized vertices~\cite{Curtis:1990zs},
are important.

\begin{acknowledgments}
\end{acknowledgments}
The author thanks to W.-S. Hou, T. Inagaki, Y. Mimura, M. Kohda
and H. Mineo for discussions.
The author is supported by Ministry of Science and Technology
(Taiwan, ROC), through Grant No. MOST 103-2811-M-002-087.


\end{document}